\documentclass[twocolumn]{aastex62}

\begin{document}
\title{\bf \large Predicting Quasar Continua Near Lyman-$\alpha$ with Principal Component Analysis}

\author{Frederick B. Davies}
\affiliation{Department of Physics, University of California, Santa Barbara, CA 93106-9530, USA}

\author{Joseph F. Hennawi}
\affiliation{Department of Physics, University of California, Santa Barbara, CA 93106-9530, USA}

\author{Eduardo Ba\~{n}ados}
\altaffiliation{Carnegie-Princeton Fellow}
\affiliation{The Observatories of the Carnegie Institution for Science, 813 Santa Barbara Street, Pasadena, California 91101, USA}

\author{Robert A. Simcoe}
\affiliation{MIT-Kavli Center for Astrophysics and Space Research, 77 Massachusetts Avenue, Cambridge, Massachusetts 02139, USA}

\author{Roberto Decarli}
\affiliation{Max Planck Institut f\"{u}r Astronomie, K\"{o}nigstuhl 17, D-69117 Heidelberg, Germany}
\affiliation{INAF--Osservatorio Astronomico di Bologna, via Gobetti 93/3, 40129 Bologna, Italy}

\author{Xiaohui Fan}
\affiliation{Steward Observatory, The University of Arizona, 933 North Cherry Avenue, Tucson, Arizona 85721-0065, USA}

\author{Emanuele P. Farina}
\affiliation{Max Planck Institut f\"{u}r Astronomie, K\"{o}nigstuhl 17, D-69117 Heidelberg, Germany}
\affiliation{Department of Physics, University of California, Santa Barbara, CA 93106-9530, USA}

\author{Chiara Mazzucchelli}
\affiliation{Max Planck Institut f\"{u}r Astronomie, K\"{o}nigstuhl 17, D-69117 Heidelberg, Germany}

\author{Hans-Walter Rix}
\affiliation{Max Planck Institut f\"{u}r Astronomie, K\"{o}nigstuhl 17, D-69117 Heidelberg, Germany}

\author{Bram P. Venemans}
\affiliation{Max Planck Institut f\"{u}r Astronomie, K\"{o}nigstuhl 17, D-69117 Heidelberg, Germany}

\author{Fabian Walter}
\affiliation{Max Planck Institut f\"{u}r Astronomie, K\"{o}nigstuhl 17, D-69117 Heidelberg, Germany}

\author{Feige Wang}
\affiliation{Department of Astronomy, School of Physics, Peking University, Beijing 100871, China}
\affiliation{Kavli Institute for Astronomy and Astrophysics, Peking University, Beijing 100871, China}
\affiliation{Department of Physics, University of California, Santa Barbara, CA 93106-9530, USA}

\author{Jinyi Yang}
\affiliation{Department of Astronomy, School of Physics, Peking University, Beijing 100871, China}
\affiliation{Kavli Institute for Astronomy and Astrophysics, Peking University, Beijing 100871, China}
\affiliation{Steward Observatory, The University of Arizona, 933 North Cherry Avenue, Tucson, Arizona 85721-0065, USA}

\email{davies@physics.ucsb.edu}

\begin{abstract}
Measuring the proximity effect and the damping wing of intergalactic neutral hydrogen in quasar spectra during the epoch of reionization requires an estimate of the intrinsic continuum at rest-frame wavelengths $\lambda_{\rm rest}\sim1200$--$1260$ {\AA}. In contrast to previous works which used composite spectra with matched spectral properties or explored correlations between parameters of broad emission lines, we opted for a non-parametric predictive approach based on principal component analysis (PCA) to predict the intrinsic spectrum from the spectral properties at redder (i.e. unabsorbed) wavelengths. We decomposed a sample of $12764$ spectra of $z\sim2$--$2.5$ quasars from SDSS/BOSS into 10 red-side ($1280$ {\AA} $<\lambda_{\rm rest}<2900$ {\AA}) and 6 blue-side ($1180$ {\AA} $<\lambda_{\rm rest}<1280$ {\AA}) PCA basis spectra, and constructed a projection matrix to predict the blue-side coefficients from a fit to the red-side spectrum. We found that our method predicts the blue-side continuum with $\sim6$--$12\%$ precision and $\la1\%$ bias by testing on the full training set sample. We then computed predictions for the blue-side continua of the two quasars currently known at $z>7$: ULAS J1120+0641 ($z=7.09$) and ULAS J1342+0928 ($z=7.54$). Both of these quasars are known to exhibit extreme emission line properties, so we individually calibrated the precision of the continuum predictions from similar quasars in the training set. We find that both $z>7$ quasars, and in particular ULAS J1342+0928, show signs of damping wing-like absorption at wavelengths redward of Ly$\alpha$.
\end{abstract}

\section{Introduction}

The damping wing of neutral hydrogen Ly$\alpha$ absorption in the intergalactic medium (IGM) is predicted to be a key signature of the epoch of reionization at $z>6$ \citep{ME98}. This damped absorption signature should be very broad, affecting rest-frame wavelengths redward of Ly$\alpha$ ($\lambda_{\rm rest}=1215.67$ {\AA}) out to $\lambda_{\rm rest}\sim1260$ {\AA} if the IGM is mostly neutral.  Measurement of this signal, however, requires knowledge of the \emph{intrinsic} (i.e. unabsorbed) profile of the quasar spectrum, which in the wavelength range relevant to the IGM damping wing consists of a combination of multiple Ly$\alpha$ and \ion{N}{5} broad emission line components in addition to a smooth underlying continuum. Our ability to measure the IGM damping wing is thus mostly limited by our ability to predict the shape of this part of the quasar spectrum. 

Lacking an accurate physical model to predict the combined emission from the quasar accretion disk and the broad-line region, we must instead resort to an empirical approach, perhaps aided by machine learning. Fortunately, thousands of quasars have been observed by SDSS, and these spectra hold a wealth of information relating the correlated strengths and properties of their broad emission lines. The challenge lies in how exactly to extract these correlations quantitatively to predict one part of the spectrum from measurements of a different part. In this case, we would like to predict the spectral region potentially affected by IGM absorption ($\lambda_{\rm rest}<1280$ {\AA}, henceforth the ``blue-side" of the spectrum) from the remaining redward spectral coverage ($\lambda_{\rm rest}>1280$ {\AA}, henceforth the ``red-side" of the spectrum).

Correlations between various spectral features in the spectra of quasars have been studied for decades (e.g. \citealt{BG92}), and strong correlations are known to exist between various broad emission lines from the rest-frame ultraviolet to the optical (e.g. \citealt{Shang07}). In principle, then, it should be possible to use the information contained in the red-side portion of the quasar spectrum to predict the quasar continuum on the blue-side. Various techniques exist in the literature for predicting the quasar continuum close to Ly$\alpha$, including the direct approach of Gaussian fitting the red side of the Ly$\alpha$ line (e.g. \citealt{KH09}) and stacking of quasar spectra with similar (non-Ly$\alpha$) emission line properties (e.g. \citealt{Mortlock11,Simcoe12,Banados17}). The most sophisticated predictive model to date was presented by \citet{Greig17a}, who determined covariant relationships between parameters of Gaussian fits to broad emission lines of \ion{C}{4}, \ion{C}{3}], and \ion{S}{4}+\ion{O}{4}] and those of Gaussian fits to the Ly$\alpha$ line.

Complicating matters is the fact that the spectral properties of quasars at $z\ga6.5$ -- most notably the properties of their \ion{C}{4} broad emission lines -- appear to preferentially occupy a sparsely populated tail in the distribution of lower-redshift quasars \citep{Mazzucchelli17}. The two quasars known at $z>7$, and in particular the newly discovered highest-redshift quasar ULAS J1342+0928 \citep{Banados17}, show very large \ion{C}{4} blueshifts relative to the general quasar population. The blueshift of the \ion{C}{4} emission line is correlated with properties of the Ly$\alpha$ emission line (e.g. \citealt{Richards11}), and thus must be properly accounted for when modeling the Ly$\alpha$ region of high-redshift quasars \citep{BB15}. Any method trained on typical low-redshift quasars must then be able to perform well on objects which lie in the tails of the distribution of spectral properties\footnote{Another option would be to simply restrict the training set to quasars with similar red-side properties, however there may be too few such analogs (e.g. the 46 \ion{C}{4}-based analogs to ULAS J1342+0928 found by \citet{Banados17}) to build a flexible model.}.

A common non-parametric method for determining correlations between different regions of quasar spectra is Principal Component Analysis (PCA), wherein a set of input spectra is decomposed into eigenspectra that correspond to modes of common variations between different spectra. PCA was first applied to the spectral properties of quasars by \citet{BG92} through analysis of not the spectral pixels themselves but of the properties of various emission lines in the rest-frame optical. \citet{Francis92} was the first work to apply PCA to the spectral pixels themselves, using a relatively small sample of rest-frame UV quasar spectra to investigate PCA as a tool for quasar classification (see also \citealt{Yip04,Suzuki06}). The idea of using PCA to predict the intrinsic continuum of absorbed regions of the spectrum was introduced by \citet{Suzuki05}, who constructed a predictive PCA model from low-redshift ($z\sim0.14$--$1.04$) quasars in the context of predicting the unabsorbed continuum in the Ly$\alpha$ forest of higher-redshift quasars where the continuum level cannot be measured directly. This technique was revisited by \citet{Paris11} with a somewhat larger sample of high signal-to-noise spectra of $z\sim3$ quasars to estimate the evolution of the Ly$\alpha$ forest mean flux. As an example application of these methods, \citet{Eilers17} recently applied the PCA models of \citet{Suzuki05} and \citet{Paris11} to measure the sizes of proximity zones in $z\sim6$ quasar spectra.

In this work, we develop a PCA-based model for predicting the blue-side quasar continuum from the red-side spectrum in a similar vein to \citet{Suzuki05} and \citet{Paris11}. In \S~2 we construct the ``training set" of quasar spectra that serve as the foundation of our predictive model. In \S~3 we compute a set of basis spectra via a PCA decomposition of the spectra (in log space) and calibrate a relationship (i.e. a projection matrix) between red-side and blue-side PCA coefficients. In \S~4 we test the predictive power of the projections from red-side to blue-side on the training set spectra, and quantify the resulting (weak) bias and covariant uncertainty of the predicted blue-side continua. In \S~5 we apply our predictive model to the two quasars known at $z>7$, which appear to show signs of damped Ly$\alpha$ absorption from the IGM. Finally, in \S~6 we conclude with a summary and describe future applications of this continuum model.

\section{Definition of the Training Set}


\begin{figure*}
\begin{center}
\resizebox{8.8cm}{!}{\includegraphics[trim={2.0em 0.0em 5.0em 4.5em},clip]{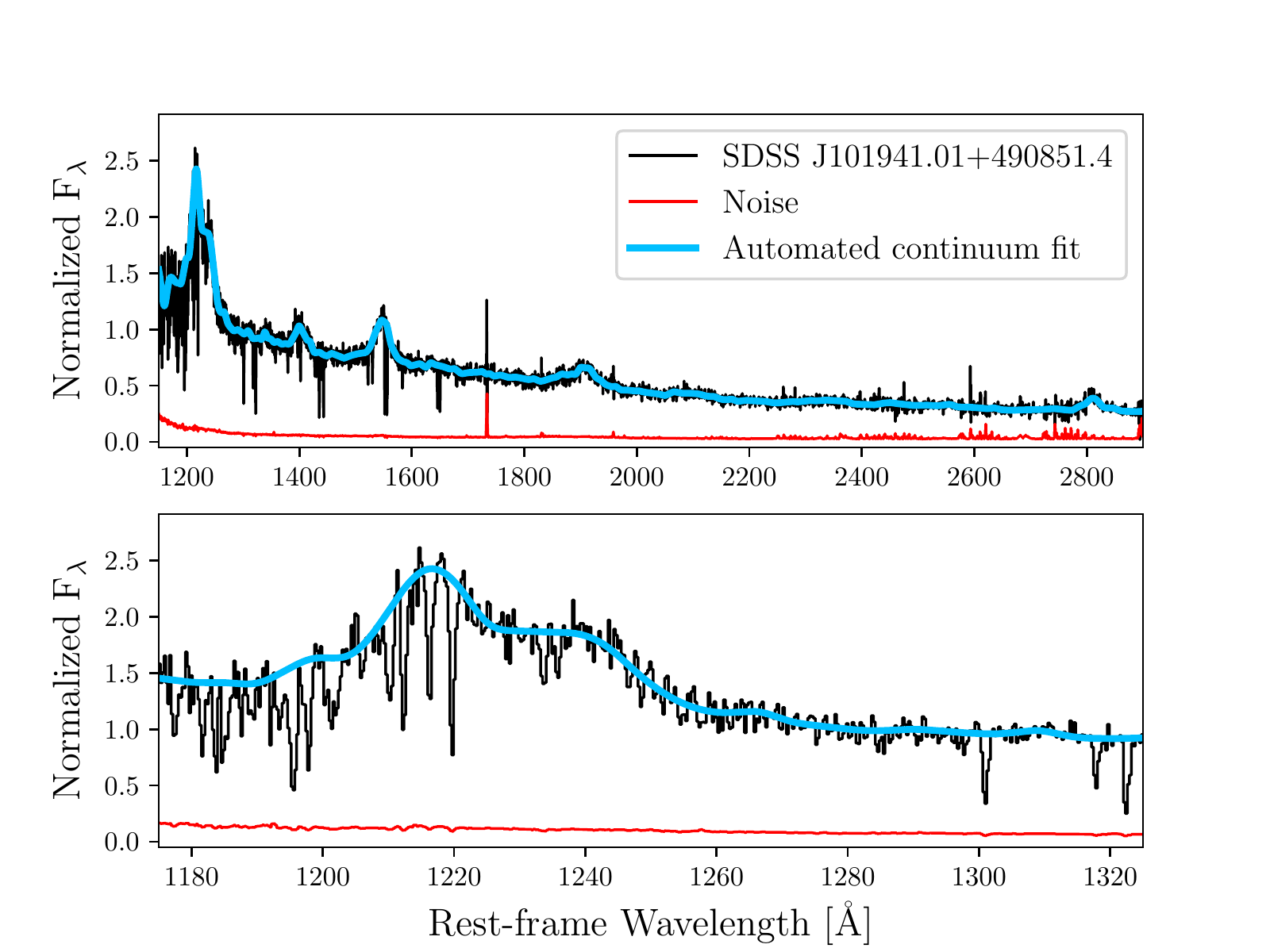}}
\resizebox{8.8cm}{!}{\includegraphics[trim={2.0em 0.0em 5.0em 4.5em},clip]{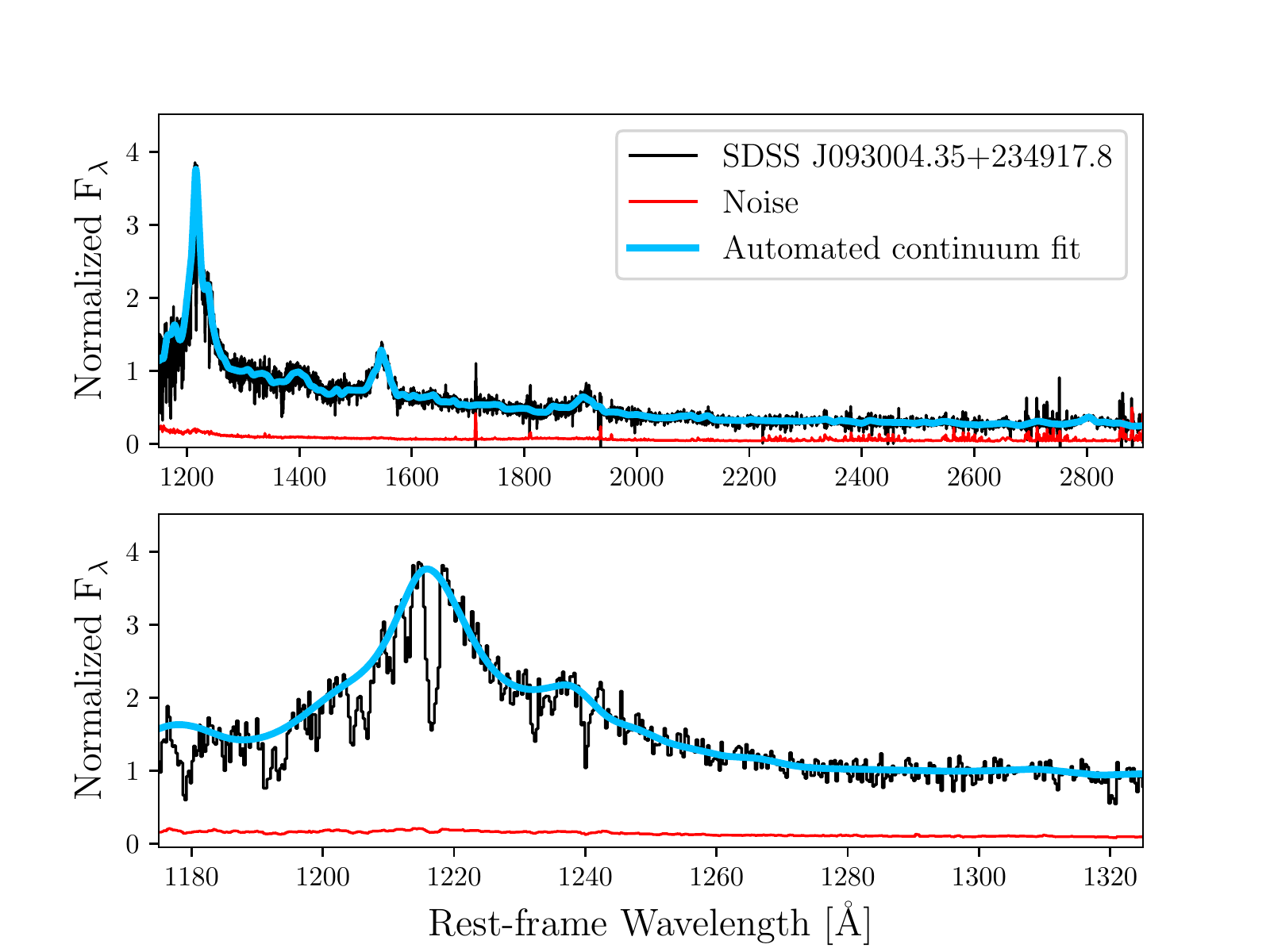}}
\end{center}
\caption{SDSS DR12 spectra of two quasars (black), their noise vectors (red), and their associated auto-fit continua (blue). The wavelength axes have been normalized to each quasar's rest-frame, and the flux densities have been normalized to unity at $\lambda_{\rm rest}=1290$ {\AA}. In addition to the Ly$\alpha$+\ion{N}{5}+\ion{Si}{2} complex at the blue end of the spectra, prominent broad emission lines of \ion{Si}{4}, \ion{C}{4}, \ion{C}{3}], and \ion{Mg}{2} are visible.}
\label{fig:cont_ex}
\end{figure*}

\begin{figure*}
\begin{center}
\resizebox{8.8cm}{!}{\includegraphics[trim={2.75em 0.0em 5.0em 4.5em},clip]{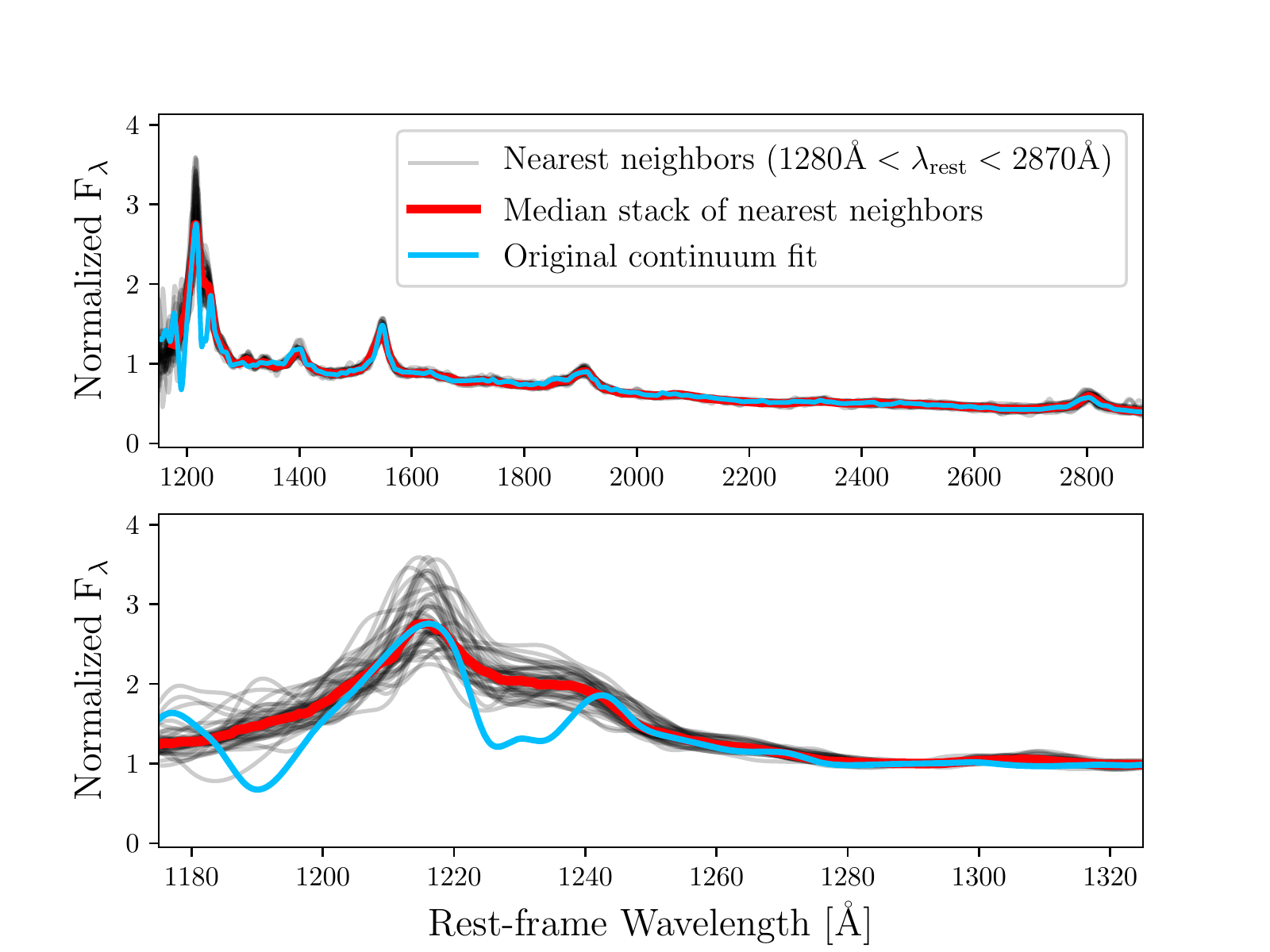}}
\resizebox{8.8cm}{!}{\includegraphics[trim={2.75em 0.0em 5.0em 4.5em},clip]{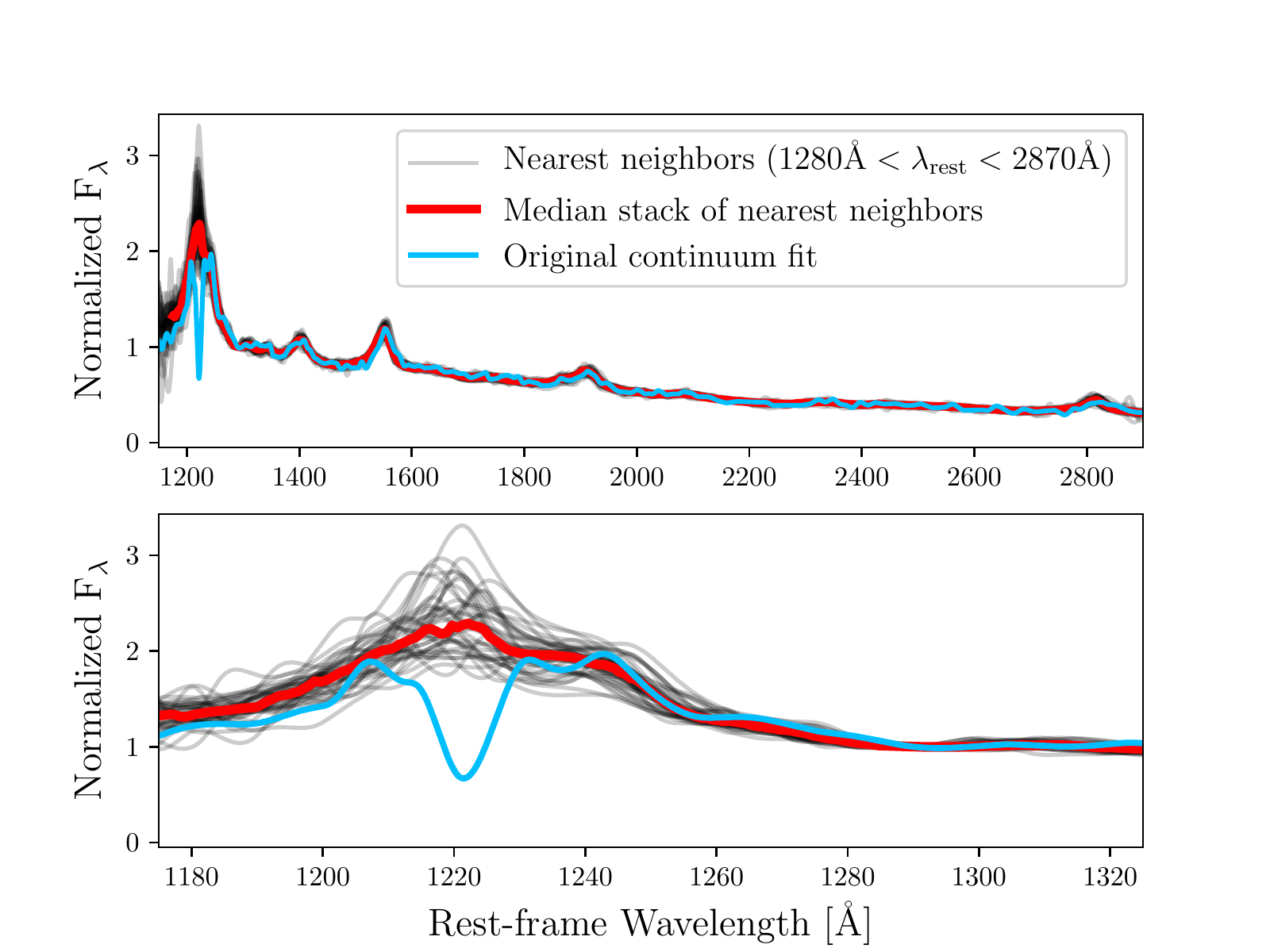}}
\end{center}
\caption{Auto-fit continua (blue) of SDSS J000113.15+322331.8 (left) and SDSS J235144.37+085649.4 (right), their 40 nearest neighbor spectra for wavelengths $1280 {\rm \AA} < \lambda_{\rm rest} < 2870 {\rm \AA}$ (grey), and the median stacks of those nearest neighbors (red). The strong associated absorption in the original spectra, shown more clearly in the bottom panels, is no longer present in the nearest neighbor stacks.}
\label{fig:nnstack_ex}
\end{figure*}

We draw our training set of quasar spectra from SDSS-III/BOSS \citep{Eisenstein11,Dawson13}, which obtained $R\sim1200$--$2500$ spectra covering $\lambda_{\rm obs}\sim3560$--$10400$ {\AA} \citep{Smee13} of $294,512$ quasars. We make use of the SDSS DR12 \citep{Alam15} quasar catalog \citep{Paris17} with a query to the quasar spectrum database IGMSpec \citep{Prochaska17} for quasars with BOSS pipeline redshifts $2.09<z_{\rm pipe}<2.51$\footnote{Motivated by the Appendix of \citet{Greig17a}, we use $z_{\rm pipe}$ rather than $z_{\rm PCA}$ because the \ion{C}{4} emission line shifts do not show redshift dependencies due to sky lines.}, \texttt{BAL\_FLAG\_VI} $=0$ to reject broad absorption line (BAL) quasars, and \texttt{ZWARNING} $=0$ to avoid objects with highly uncertain redshifts. This redshift range was chosen for similar reasons as \citet{Greig17a}: the spectra cover the Ly$\alpha$ and \ion{Mg}{2} broad emission lines.

We then compute the median signal-to-noise at $\lambda_{\rm rest} = \lambda_{\rm obs}/(1+z_{\rm pipe}) = 1290\pm2.5$\AA\, for each spectrum, and apply a signal-to-noise cut of ${\rm S\slash N}>7.0$. This selection resulted in 13,328 quasars with complete wavelength coverage from $\lambda_{\rm rest}\sim1170$--$2900$ {\AA}, covering a similar range as observed for $z>7$ quasars, and a median signal-to-noise of 10.1 at $\lambda_{\rm rest}=1290\pm2.5$ {\AA}. The signal-to-noise ratio in this wavelength range is what we will refer to as the ``${\rm S\slash N}$" of the spectra. Each spectrum was then normalized such that its median flux at $\lambda_{\rm rest} = 1290\pm2.5$ {\AA} is unity.

We then applied an automated continuum-fitting method developed by \citet{Young79} and \citet{Carswell82}, as implemented by \citet{Dall'Aglio08}, which is designed to recover a smooth continuum in the presence of absorption lines both inside and outside of the Ly$\alpha$ forest. In brief, the method consists of dividing the spectra into 16-pixel segments ($\sim1100$ km/s), fitting continuous cubic spline functions to each segment, and iteratively rejecting pixels in each segment which lie more than two standard deviations below the fit (i.e. pixels inside of absorption lines). 

We show two examples of quasar spectra (${\rm S\slash N}\sim12$) and their continuum fits in Figure~\ref{fig:cont_ex}. At this stage, we identify spectra whose continuum fits (normalized to be unity at $\lambda_{\rm rest}=1290$ {\AA}) fall below 0.5 at $\lambda_{\rm rest}<1280$ {\AA} and remove them from the analysis -- these $357$ quasars exhibit the strongest associated absorption (e.g. damped Ly$\alpha$ absorbers, strong \ion{N}{5} absorption) or are remaining BAL quasars which were not flagged by the visual inspection of \citet{Paris17}. We additionally remove $207$ quasars whose continua drop below 0.1 at $\lambda_{\rm rest}>1280$ {\AA}, because such a weak continuum relative to $\lambda_{\rm rest}=1290$ {\AA} is indicative of very low ${\rm S\slash N}$ in the red-side spectrum, which typically implies significant systematics from OH line sky-subtraction residuals. Applying all of these criteria, our final training set consists of the remaining $12,764$ quasar spectra.

Our sample of auto-fit quasar continua still contains a small fraction of ``junk," typically quasars with strong associated absorption that were not caught by the blue-side criterion above or other artifacts that are not straightforward to eliminate in an automated fashion. To further clean up the training set in an objective manner, we replace each spectrum with a median stack of the original spectrum and its 40 nearest-neighbors in the set of auto-fit continua, where the neighbors have been defined via a Euclidean distance in (normalized) flux units. That is,
\begin{equation}
D_{ij} = \sqrt{\sum_\lambda(C_{\lambda,i}-C_{\lambda,j})^2},
\end{equation}
where $i$ and $j$ denote two different quasar spectra and $C_\lambda$ is the normalized auto-fit continuum.  To avoid combining spectra with similar associated absorbers present in the Ly$\alpha$+\ion{N}{5} region, we find the nearest-neighbors only using pixels with $\lambda_{\rm rest}>1280$ {\AA}. In Figure~\ref{fig:nnstack_ex} we show the resulting ``nearest-neighbor stacks" (red) for two ${\rm S\slash N}\sim11$ quasars with strong associated absorbers, seen as the large dips in the original continuum fit (cyan) close to Ly$\alpha$, whose spectra would otherwise contribute unwanted features (i.e. not intrinsic quasar emission) to the analysis.

\section{log-space PCA Decomposition and Blue-side Projection}\label{sec:pca}

\begin{figure*}
\begin{center}
\resizebox{8.8cm}{!}{\includegraphics[trim={3em 3.5em 3em 6em},clip]{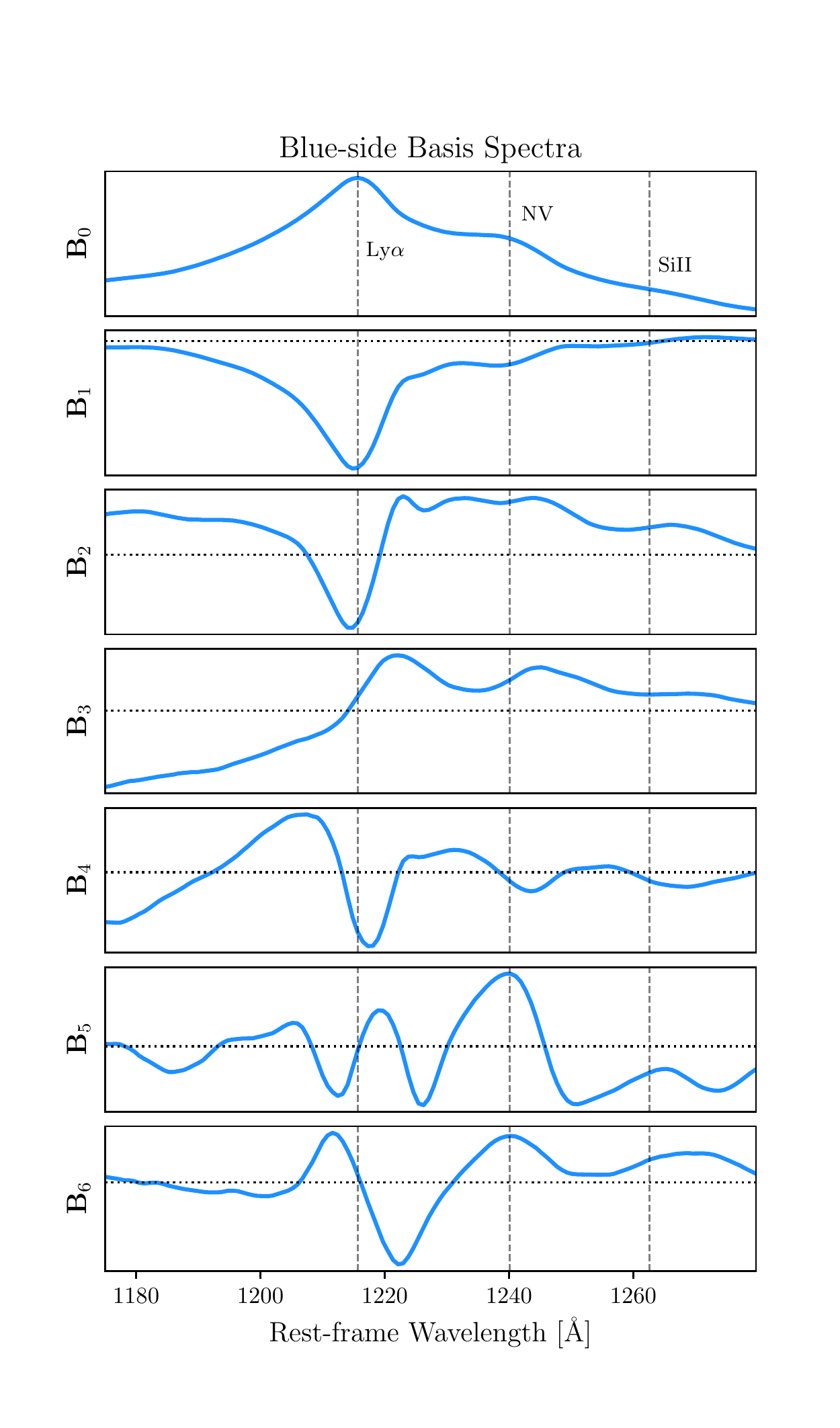}}
\resizebox{8.8cm}{!}{\includegraphics[trim={3em 3.5em 3em 6em},clip]{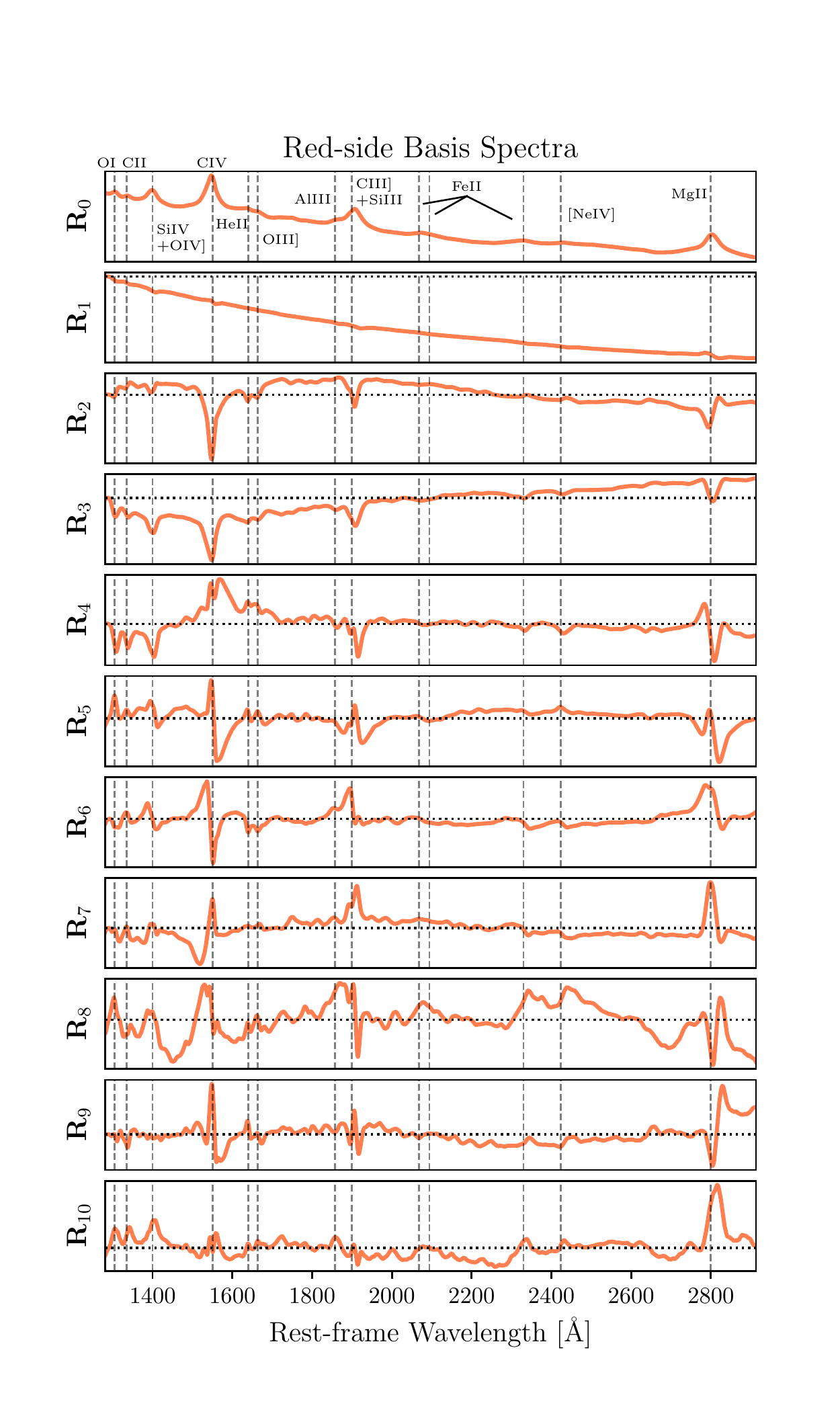}}\\
\end{center}
\caption{Blue-side and red-side basis spectra derived from the log-space PCA decomposition of 12,764 SDSS DR12 quasar spectra. The ``0th" component in the top panels represents the mean of the \emph{log} of the spectra, while the lower panels show the basis spectra ordered from highest to lowest variance explained from top to bottom. Vertical dashed lines highlight the central wavelengths of transitions of various species (or average wavelengths, in the case of blends) corresponding to broad emission lines in the spectrum, with line identifications taken from the SDSS/BOSS composite spectrum of \citet{Harris16}. The horizontal dotted line in each panel represents the zero level.}
\label{fig:pca_comps}
\end{figure*}

\begin{figure*}
\begin{center}
\resizebox{17cm}{!}{\includegraphics[trim={6.0em 2.5em 6em 4em},clip]{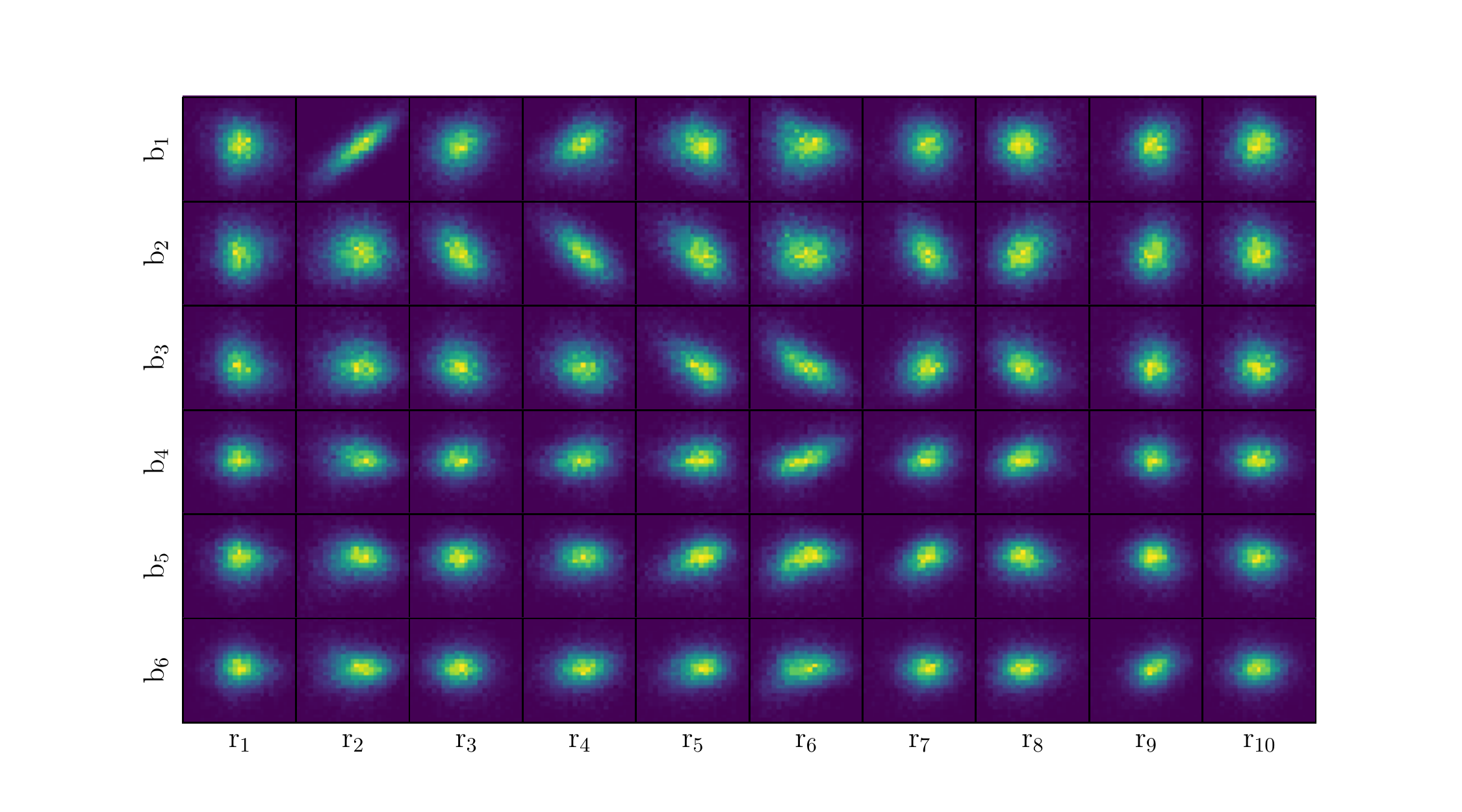}}\\
\end{center}
\caption{Joint distributions of best-fit red-side (horizontal axis) and blue-side (vertical axis) PCA coefficients from the training set. Strong and weak (anti-)correlations are visible between several of the coefficients; it is the information in these correlations that allow us to predict the blue-side spectrum from the red-side.}
\label{fig:eigen_corr}
\end{figure*}

\begin{figure*}
\begin{center}
\resizebox{8.8cm}{!}{\includegraphics[trim={0em 0.25em 4.5em 4.5em},clip]{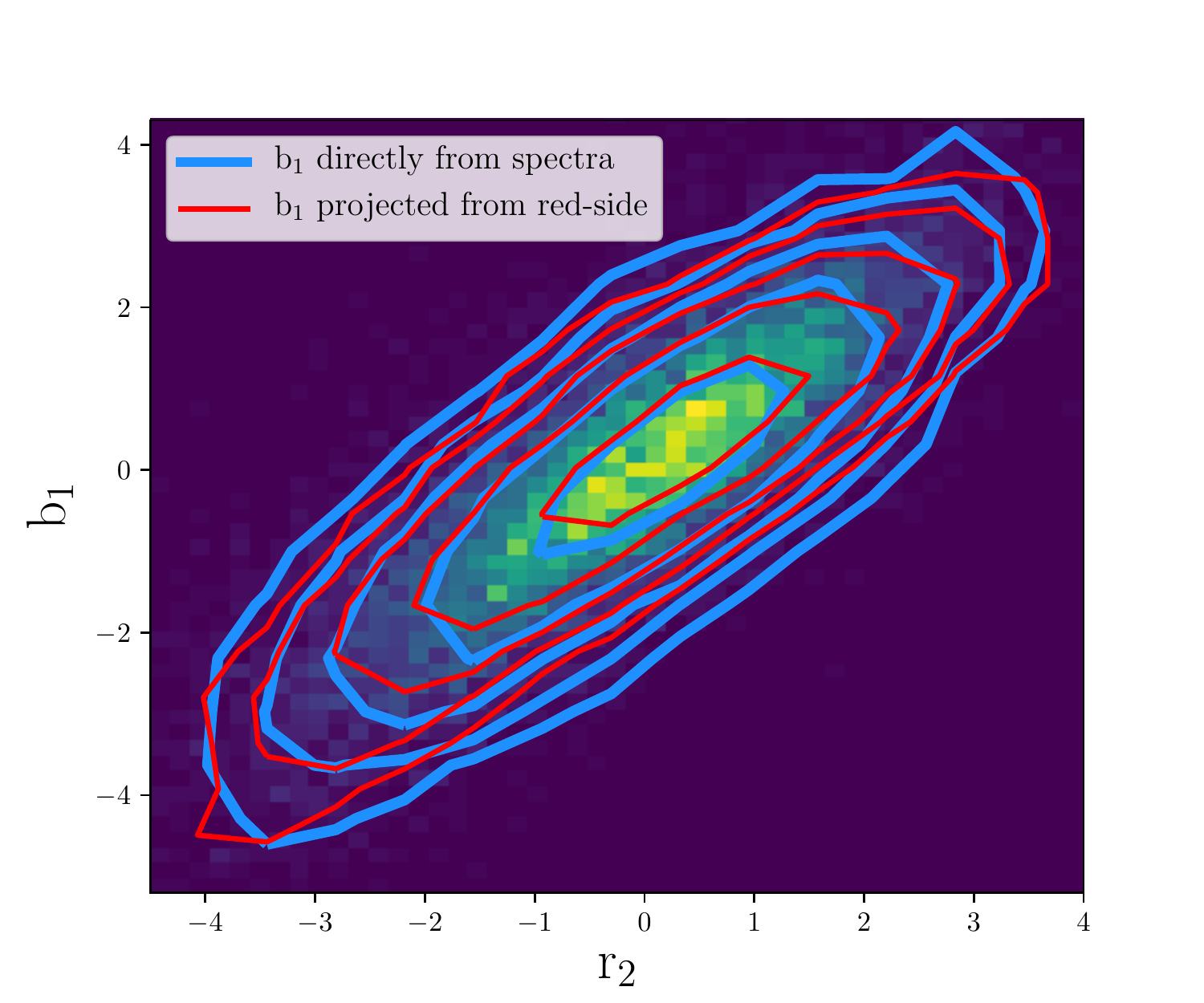}}
\resizebox{8.8cm}{!}{\includegraphics[trim={0em 0.25em 4.5em 4.5em},clip]{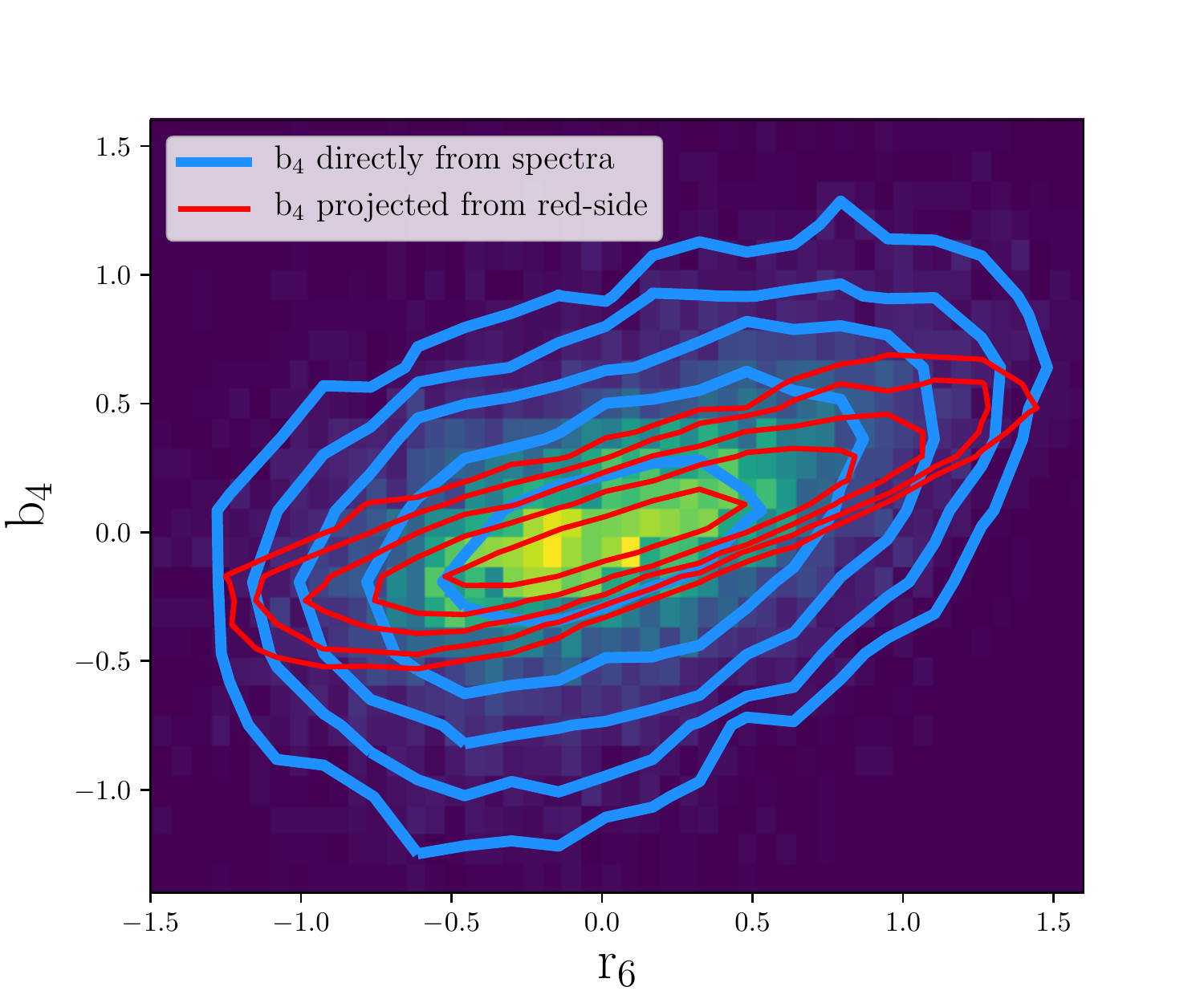}}\\
\end{center}
\caption{Left: Distribution of $r_2$ and $b_1$ coefficients determined for the nearest-neighbor-stacked training set spectra (2D histogram and blue contours), as discussed in \S~\ref{sec:pca}. The red contours show the distribution of $r_2$ and projected $b_1$ after applying equation~(\ref{eqn:proj}) to the full set of $r_i$ for each spectrum. Right: The same as the left panel, but for $r_6$ and $b_4$. The excess scatter in the blue-side coefficients determined from the spectra relative to the projections may reflect stochasticity in the relationship between the red-side and blue-side spectral features, or a non-linear relationship that is not accounted for in the projection matrix (equation \ref{eqn:proj}).}
\label{fig:projection}
\end{figure*}

PCA decomposes a set of training spectra into a set of orthogonal basis spectra, such that a spectrum can be expressed as
\begin{equation}\label{eqn:linpca}
{\bf{F}} \approx \langle {\bf F} \rangle + \sum_{i=1}^{N_{\rm PCA}} a_i {\bf{A}}_i,
\end{equation}
where $a_i$ are the coefficients associated with each basis spectrum ${\bf{A}}_i$, and $N_{\rm PCA}$ is the number of basis spectra used for the reconstruction. PCA basis spectra represent the dominant modes of variance between spectra in the training set, in order from most to least amount of variance explained. The dominant mode of variation between quasar spectra, however, is the varying slope of the (roughly) power-law continuum. Power law variations are not naturally described by a single additive component, but they are perfectly described by a single \emph{multiplicative} component, i.e. $F_\lambda=\langle F_\lambda \rangle \times \lambda^{\Delta\alpha}$ where $\langle F_\lambda \rangle$ is the average quasar spectrum and $\Delta\alpha$ is a change in spectral index. Motivated by this, we perform the PCA decomposition in \emph{log space}, 
\begin{equation}\label{eqn:logpca}
\log{\bf F} \approx \langle \log{\bf F} \rangle + \sum_{i=1}^{N_{\rm PCA}} a_i {\bf A}_i,
\end{equation}
or in other words,
\begin{equation}\label{eqn:logpca2}
{\bf F} \approx {\rm e}^{\langle \log{\bf F} \rangle} \times \prod_{i=1}^{N_{\rm PCA}} {\rm e}^{a_i {\bf A}_i}.
\end{equation}
One drawback of working in log space is that negative flux values are undefined -- fortunately, the stacked auto-fit quasar continua we are using as our training set are essentially always positive because the true continua (in the absence of artifacts) should always be positive, and as mentioned above, we have explicitly thrown out the small fraction of quasars whose auto-fit continua fall below positive thresholds.

Our goal is to predict the shape of the Ly$\alpha$ region using the information contained in the rest of the spectrum. We adopt the ``projection" procedure of \citet{Suzuki05} and \cite{Paris11}, wherein a linear mapping is constructed between the measured and predicted coefficients.
Instead of fitting the red side and projecting to a combined red+blue continuum as performed by \citet{Suzuki05} and \citet{Paris11}, we keep the red- and blue-side spectra distinct, although in practice we find that this makes little difference. We first decompose the red- and blue-side spectra with the log-space PCA described above using the PCA implementation in the python \textsc{scikit-learn} package \citep{scikit-learn}. We truncate the set of PCA basis spectra to the first 10 red-side and 6 blue-side basis spectra (${\bf R}_i$ and ${\bf B}_i$ for the red and blue sides, respectively). The choice of the number of PCA basis spectra to keep is largely arbitrary -- we chose 10 red-side basis spectra motivated by previous PCA analyses by \citet{Suzuki06} and \citet{Paris11}, and then chose the number of blue-side basis spectra such that the error in the blue-side predictions (discussed later in \S~\ref{sec:error}) did not decrease with additional spectra. Tests with increased number of red-side and blue-side basis spectra (up to 15 and 10, respectively) showed very similar results, so our analysis is not particularly sensitive to the number of basis spectra.

In Figure~\ref{fig:pca_comps} we show the red-side and blue-side mean of the log spectra (top row) and basis spectra (lower rows). Notably, the first red-side basis spectrum ${\bf R}_1$ is a smooth curve
that describes the broadband continuum variations between quasars. As mentioned above, if the variations between quasar continua were simply described by differences in spectral index (and uncorrelated with any other spectral features), the first basis spectrum should be linear in $\log{\lambda}$, and this is approximately the case.\footnote{In detail, there is a modest non-power law curvature in ${\bf R}_1$. Interpreting this curvature is beyond the scope of this work, but we note that the form of the log space decomposition is similar to those used to describe extinction curves.} The other red-side components encode correlations between the strengths of various broad emission lines and ``pseudo-continuum" features, e.g. overlapping \ion{Fe}{2} emission lines which blanket the spectrum. The blue-side components are more difficult to interpret directly, but the first two components appear to show either correlated (${\bf B}_1$) or anti-correlated (${\bf B}_2$) Ly$\alpha$ and \ion{N}{5} line strengths.

\begin{figure*}[htb]
\begin{center}
\resizebox{8.8cm}{!}{\includegraphics[trim={2.0em 0.3em 5.0em 4.4em},clip]{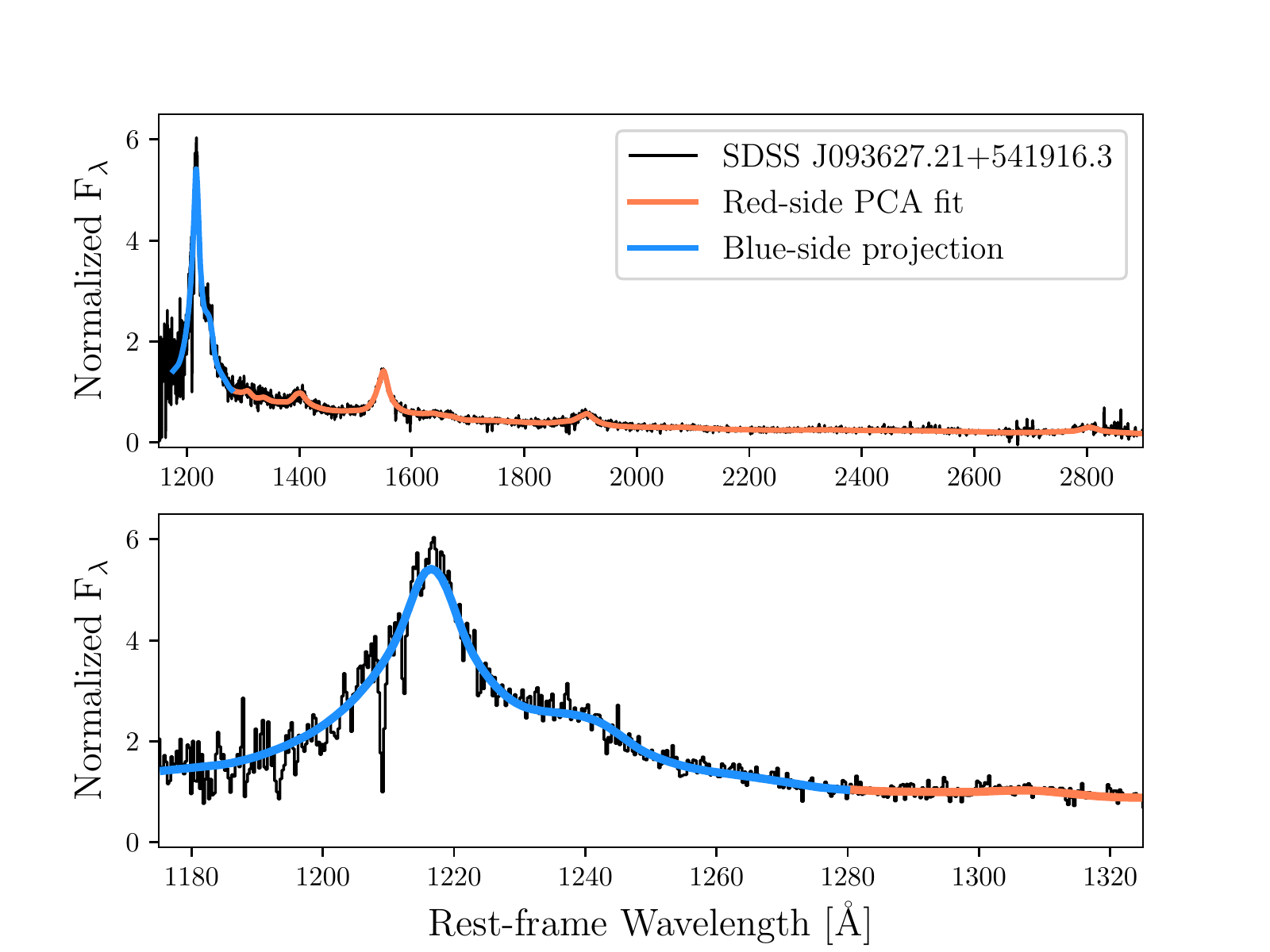}}
\resizebox{8.8cm}{!}{\includegraphics[trim={2.0em 0.3em 5.0em 4.4em},clip]{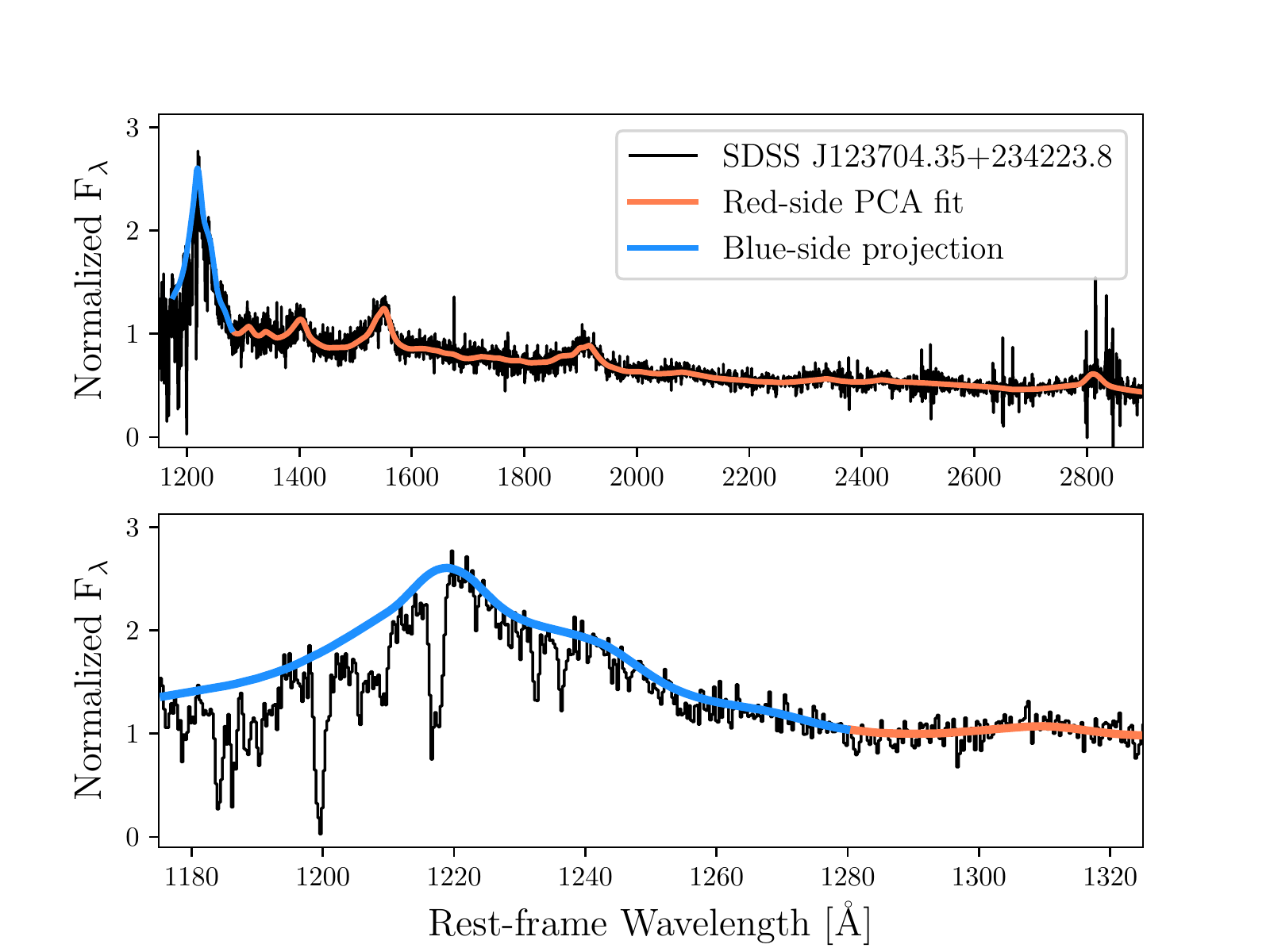}}
\end{center}
\caption{Example red-side PCA fits (orange) and blue-side predictions (blue) of two SDSS/BOSS quasar spectra (black) with ${\rm S\slash N}\sim10$ at $\lambda_{\rm rest}=1290$ {\AA}. The top panels show the entire spectrum, while the bottom panels focus on the Ly$\alpha$ region.}
\label{fig:pca_ex}
\end{figure*}

\begin{figure*}[htb]
\begin{center}
\resizebox{8.8cm}{!}{\includegraphics[trim={2.0em 0.3em 5.0em 4.4em},clip]{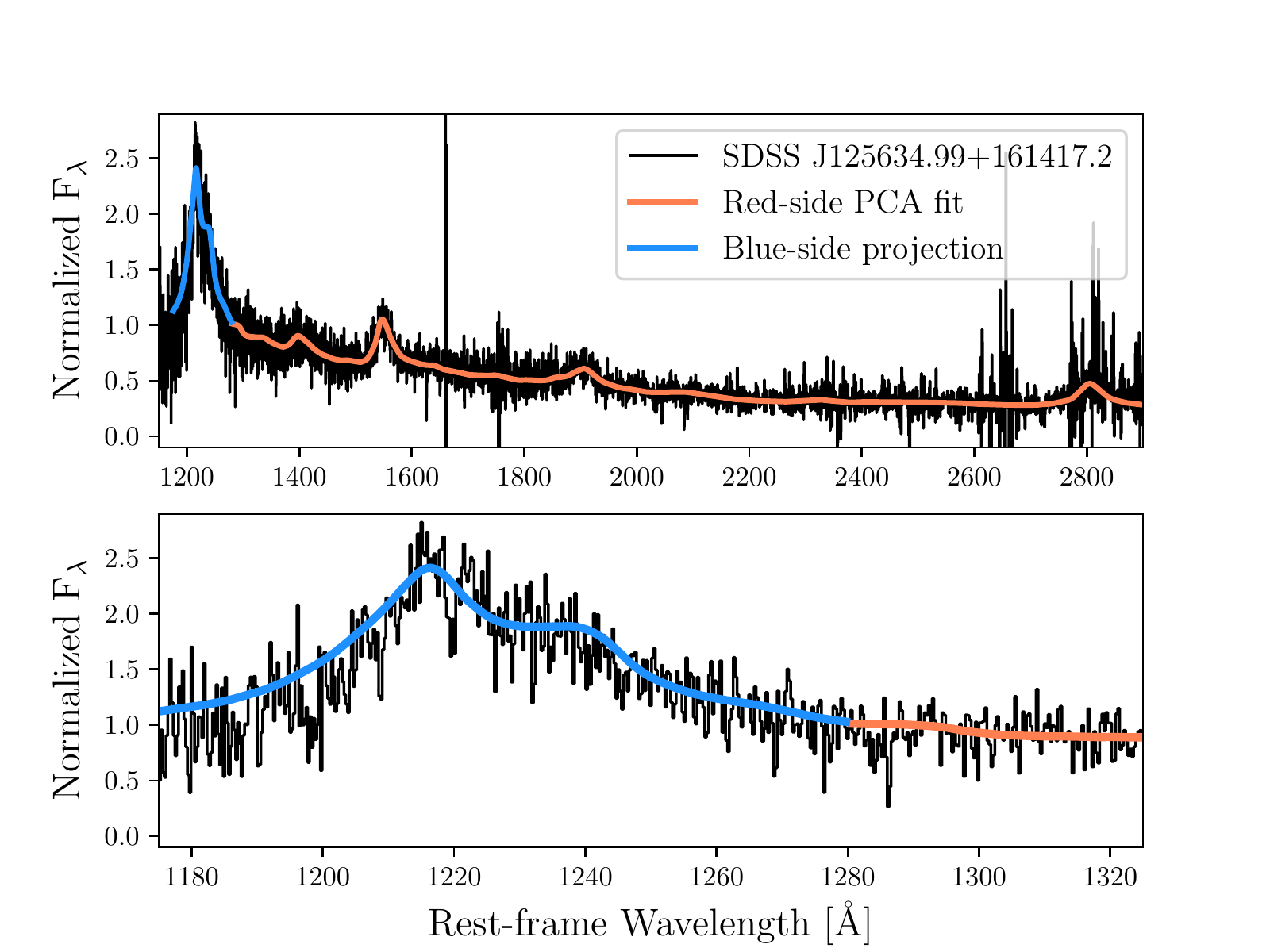}}
\resizebox{8.8cm}{!}{\includegraphics[trim={2.0em 0.3em 5.0em 4.4em},clip]{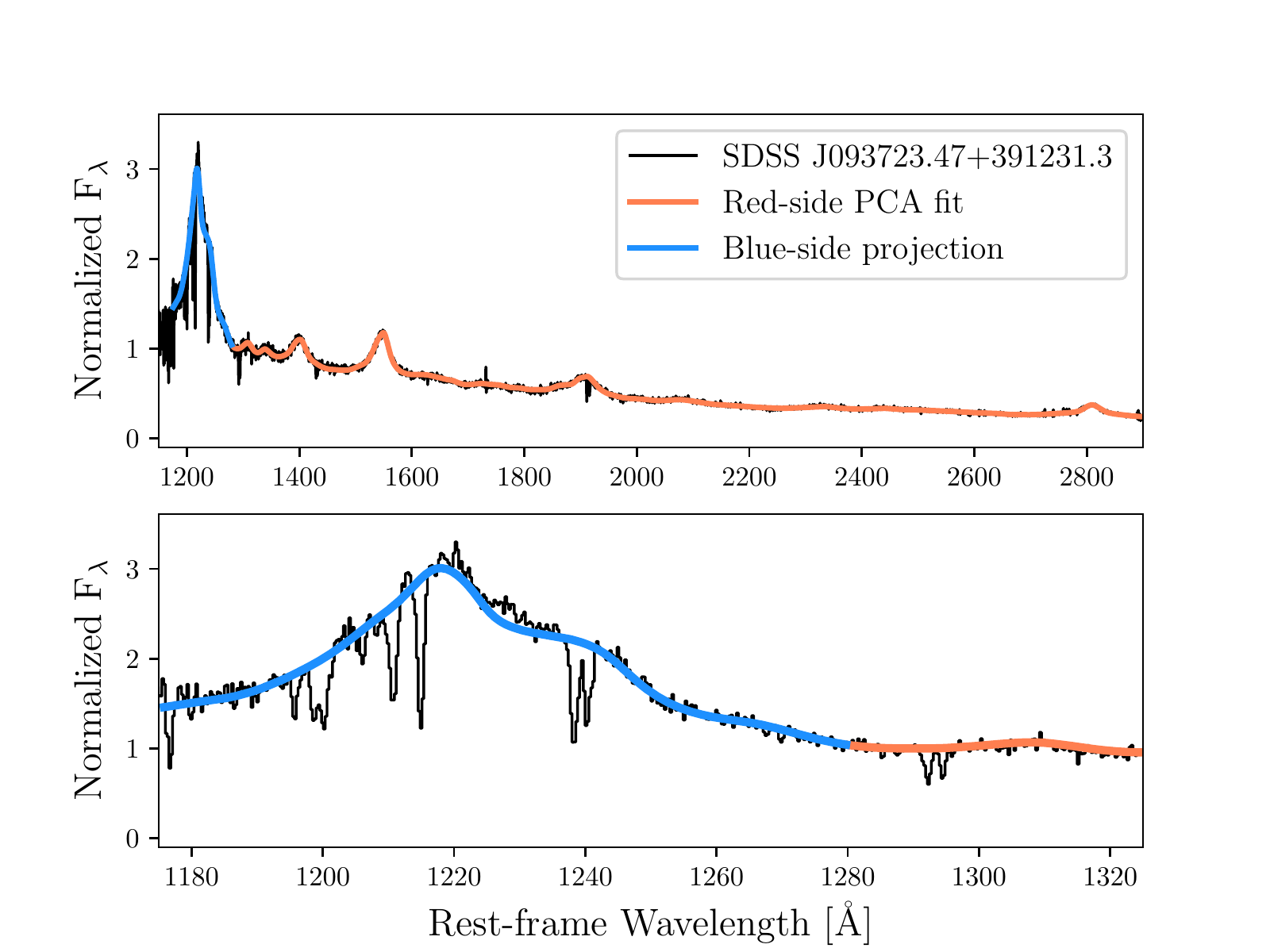}}
\end{center}
\caption{Same as Figure~\ref{fig:pca_ex} but for example quasar spectra with ${\rm S\slash N}\sim7$ (left) and ${\rm S\slash N}\sim25$ (right) at $\lambda_{\rm rest}=1290$ {\AA}.}
\label{fig:pca_ex_sn}
\end{figure*}

\begin{figure*}[htb]
\begin{center}
\resizebox{8.8cm}{!}{\includegraphics[trim={2.0em 0.3em 5.0em 4.4em},clip]{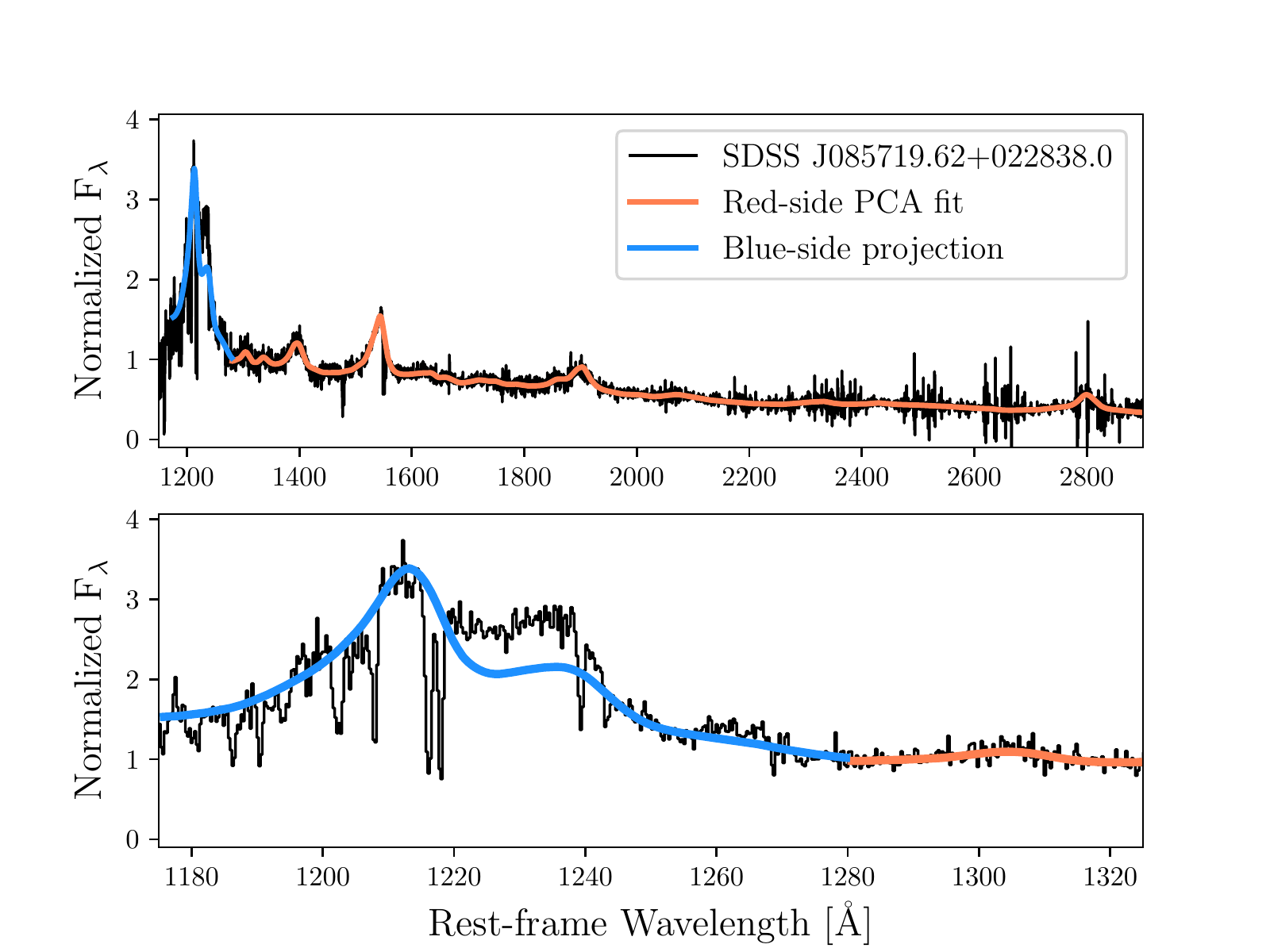}}
\resizebox{8.8cm}{!}{\includegraphics[trim={2.0em 0.3em 5.0em 4.4em},clip]{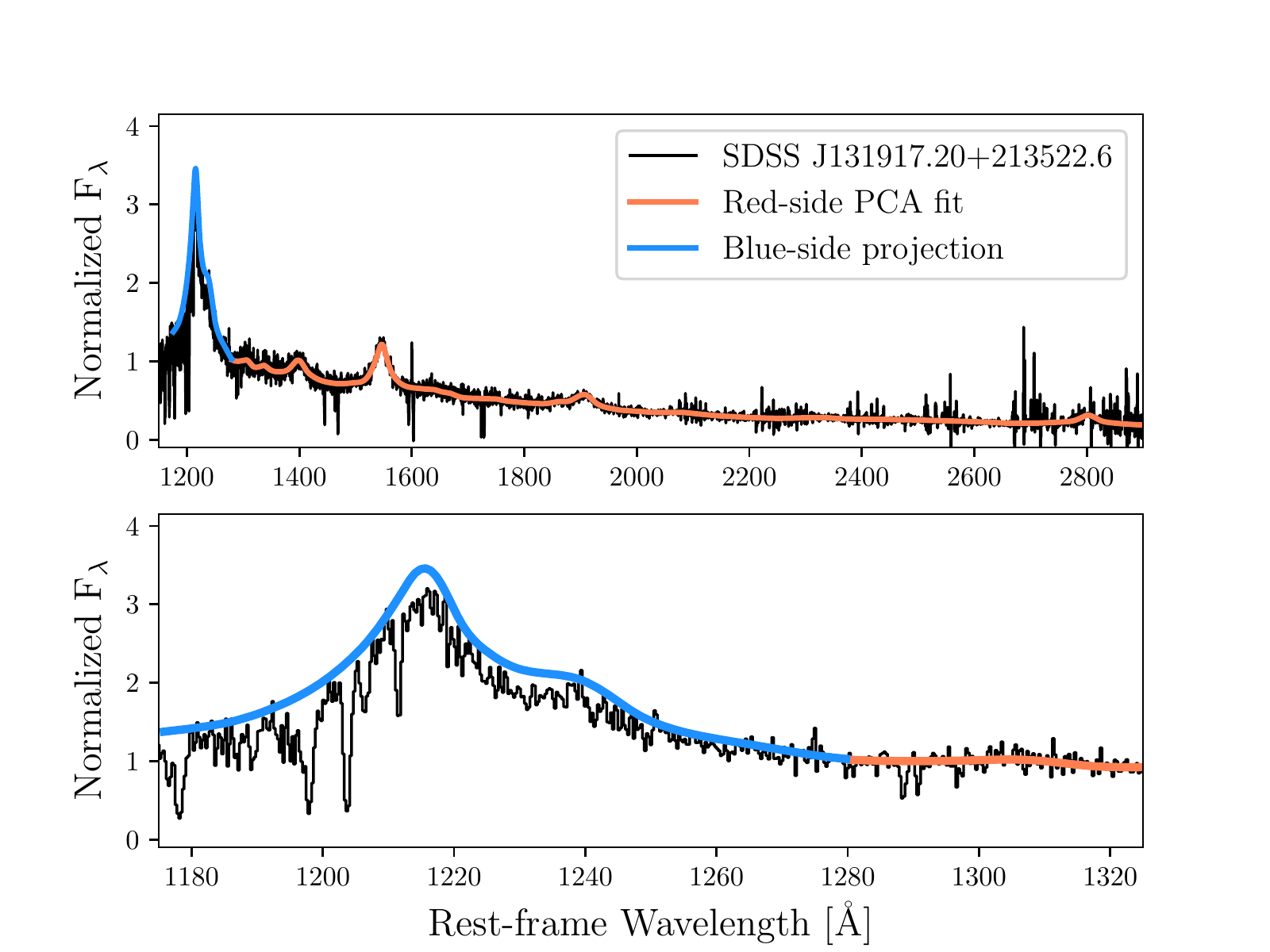}}
\end{center}
\caption{Same as Figure~\ref{fig:pca_ex} but for quasar spectra with relatively poor blue-side predictions. In the left panel we show a prediction which undershoots the true continuum, and in the right panel we show a prediction which overshoots the true continuum.}
\label{fig:pca_ex_bad}
\end{figure*}

To determine the relationship between the red-side and blue-side PCA coefficients ($r_i$ and $b_i$, respectively), we now compute the coefficients for each \emph{individual} (i.e. not median stacked) quasar spectrum in our training set of 12,764. We first fit for the red-side coefficients, $r_i$, via $\chi^2$ minimization on the \emph{original} (i.e. not auto-fit) spectra, after masking pixels which deviate more than $3\sigma$ from the auto-fit continua to remove metal absorption lines such as \ion{C}{4} and \ion{Mg}{2}. Motivated by the large velocity shifts between the systemic frame (given by, e.g., [\ion{C}{2}] emission from the host galaxy) and the broad emission lines in $z\ga6.5$ quasars (e.g. \citealt{Mazzucchelli17}), we simultaneously fit for a \emph{template redshift}, i.e. we allow the effective redshift of the PCA basis spectra be a free parameter in the fit. This template redshift, $z_{\rm temp}$, can be consistently measured for any quasar spectrum with similar spectral coverage, allowing for direct comparison between our low-redshift training set and the high-redshift quasars we are interested in, and we ascribe no physical interpretation to its value. Then, working in the $z_{\rm temp}$ frame as defined by the red-side fit, we fit the blue-side coefficients, $b_i$, via $\chi^2$ minimization on the \emph{auto-fit} spectra instead of directly from the spectral pixels because the actual blue-side spectrum is guaranteed to be contaminated by Ly$\alpha$ forest absorption at $\lambda_{\rm rest}<\lambda_{\rm Ly\alpha}$.

The predictive power of the PCA decomposition lies in the correlations between $r_i$ and $b_i$, shown in Figure~\ref{fig:eigen_corr}, which shows the joint distribution of best-fit coefficients of the training set spectra. In the left panel of Figure~\ref{fig:projection}, the 2D histogram and corresponding blue contours show the relationship between $r_2$ and $b_1$, which appears to reflect a correlation between the strength of \ion{C}{4} (red-side) and Ly$\alpha$ (blue-side) emission line strengths (see Figure~\ref{fig:pca_comps}).
Following \citet{Suzuki05} and \citet{Paris11}, we model these correlations between eigenvalues with a linear relationship, i.e. $b_i \approx \sum_{j=1}^{N_{\rm PCA,r}} r_j X_{ji}$, where ${\bf X}$ is the $N_{\rm PCA,r} \times N_{\rm PCA,b}$ projection matrix determined by solving the linear set of equations,
\begin{equation}\label{eqn:proj}
{\bf b} = {\bf r} \cdot {\bf X},
\end{equation}
where ${\bf b}$ ($N_{\rm spec}\times N_{\rm PCA,b}$) and {\bf r} ($N_{\rm spec}\times N_{\rm PCA,r}$) are the
sets of all best-fit blue-side and red-side coefficients from the training set, respectively. We solved for ${\bf X}$ using the least-squares solver in the python package \texttt{numpy} \citep{numpy}. The red contours in Figure~\ref{fig:projection} show the distribution of \emph{predicted} $b_1$ (left) and $b_4$ (right) as a function of $r_2$ and $r_6$, respectively, after application of equation~(\ref{eqn:proj}) to the full set of $r_i$. While the projection matrix provides a close approximation to the relationship between $b_1$ and $r_2$, the relationship between $b_4$ and $r_6$ in the training set spectra has considerably more scatter in $b_4$ than the predicted values. This ``excess" scatter may be related to either stochastic variations in the spectrum (e.g. the relationship between red-side and blue-side emission line properties is not exactly 1-to-1) or non-linearities in the relationship between the blue-side and red-side components that are not captured by the linear model of equation~(\ref{eqn:proj}).

We show two examples of red-side PCA fits to quasars with ${\rm S\slash N}$ close to the median of our training set and their respective blue-side predictions in Figure~\ref{fig:pca_ex}. The information contained in the shapes and amplitude of the red-side emission lines is translated into the predicted blue-side spectrum through the mapping described by equation~(\ref{eqn:proj}), and for these quasars those predictions appear to be close to the true continuum. The quality of the red-side fits and blue-side predictions are only modestly affected by ${\rm S\slash N}$, at least for the spectra selected with our ${\rm S\slash N}>7$ cut, and in Figure~\ref{fig:pca_ex_sn} we show example fits to quasars at the lower and upper ends of the distribution of ${\rm S\slash N}$ on the left (${\rm S\slash N}\sim7$) and right (${\rm S\slash N}\sim25$), respectively. To avoid only showing ``good" examples, and in a sense of full disclosure, we show two examples of particularly bad blue-side predictions\footnote{These poor predictions were discovered by inspecting spectra whose residuals at $\lambda_{\rm rest}\sim1230$ {\AA} were at the upper and lower ends of the distribution.} in Figure~\ref{fig:pca_ex_bad}. We quantify the general accuracy and precision of the blue-side predictions below.

\begin{figure*}[htb]
\begin{center}
\resizebox{8.80cm}{!}{\includegraphics[trim={1.0em 0.3em 0.0em 0.5em}]{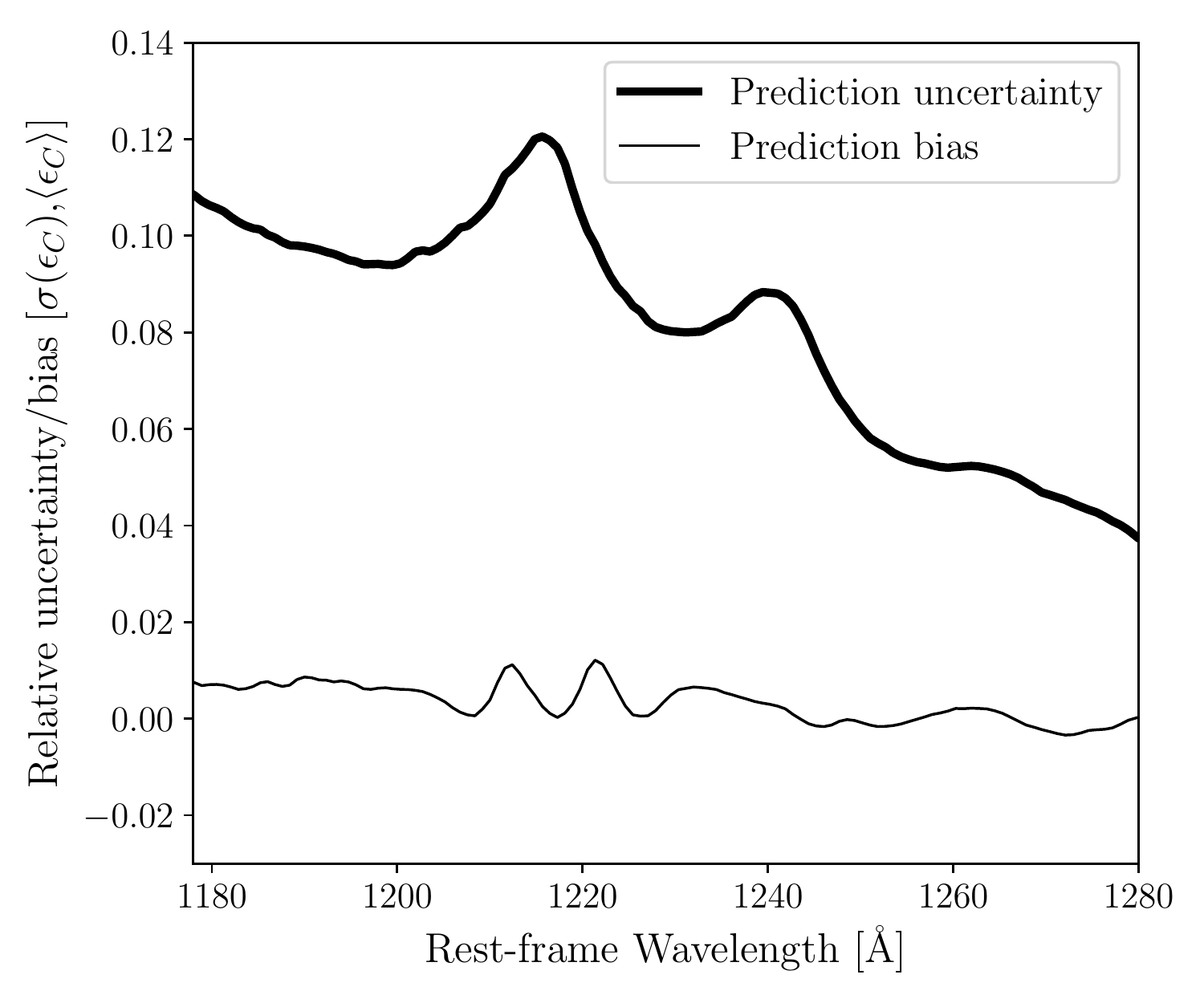}}
\resizebox{8.80cm}{!}{\includegraphics[trim={1.0em 1.0em 5.2em 0}]{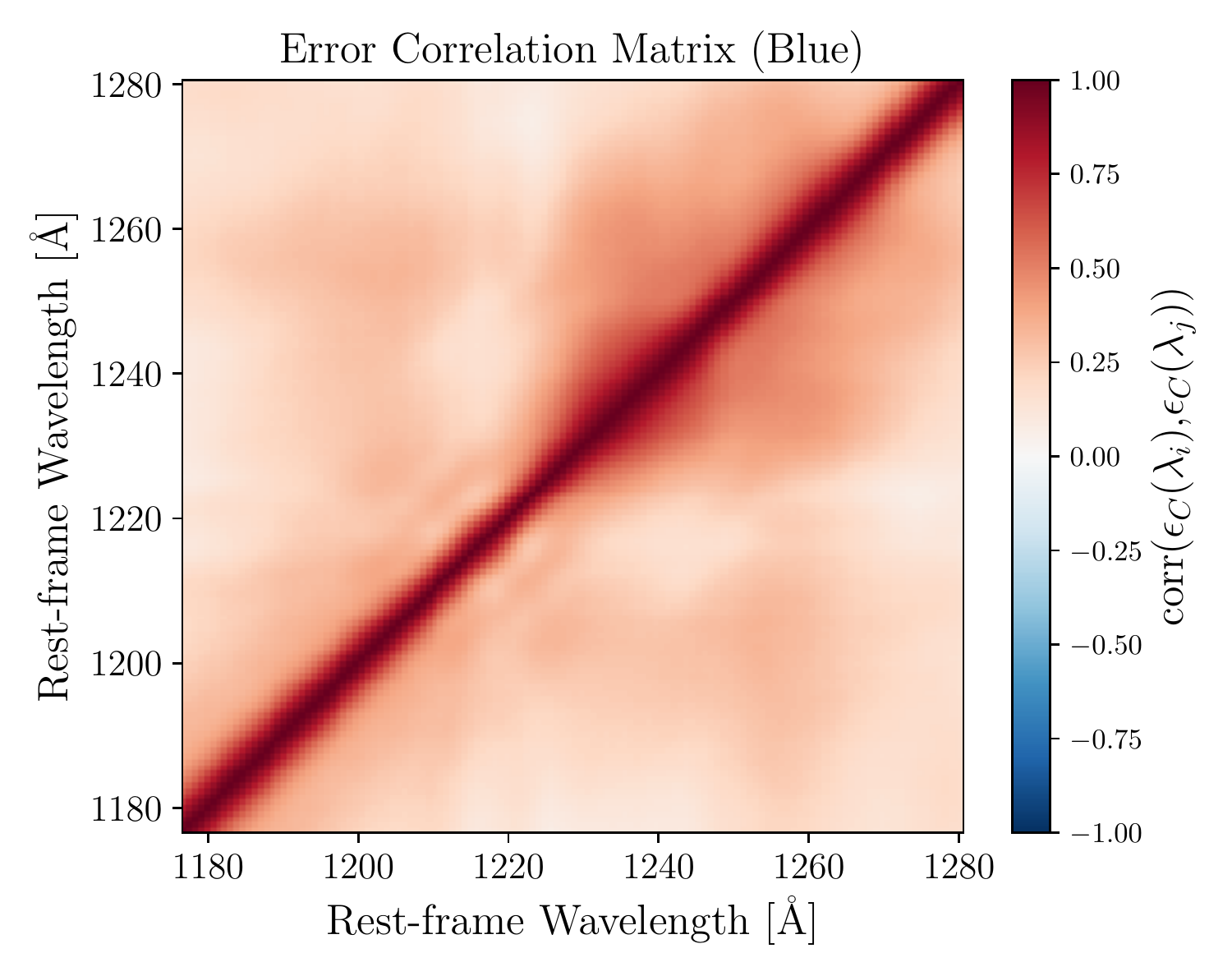}}
\end{center}
\caption{Left: Relative uncertainty ($1\sigma$, thick curve) and mean bias (thin curve) of the blue-side PCA predictions applied to the full training set sample. Locations of broad emission lines of Ly$\alpha$ and \ion{N}{5} appear as spectral regions with increased uncertainty. The most important wavelength range for constraints from the proximity zone and IGM damping wing is $\lambda_{\rm rest}\sim1210$--$1250$ {\AA}. Right: Correlation matrix of PCA blue-side prediction errors. The blurring along the diagonal is due to the scale of the spline fit continua, while the larger-scale correlations are due to errors in matching broad emission line features.}
\label{fig:fit_err_blue}
\end{figure*}

\begin{figure*}[htb]
\begin{center}
\resizebox{8.80cm}{!}{\includegraphics[trim={1.0em 0.3em 0.0em 0.5em}]{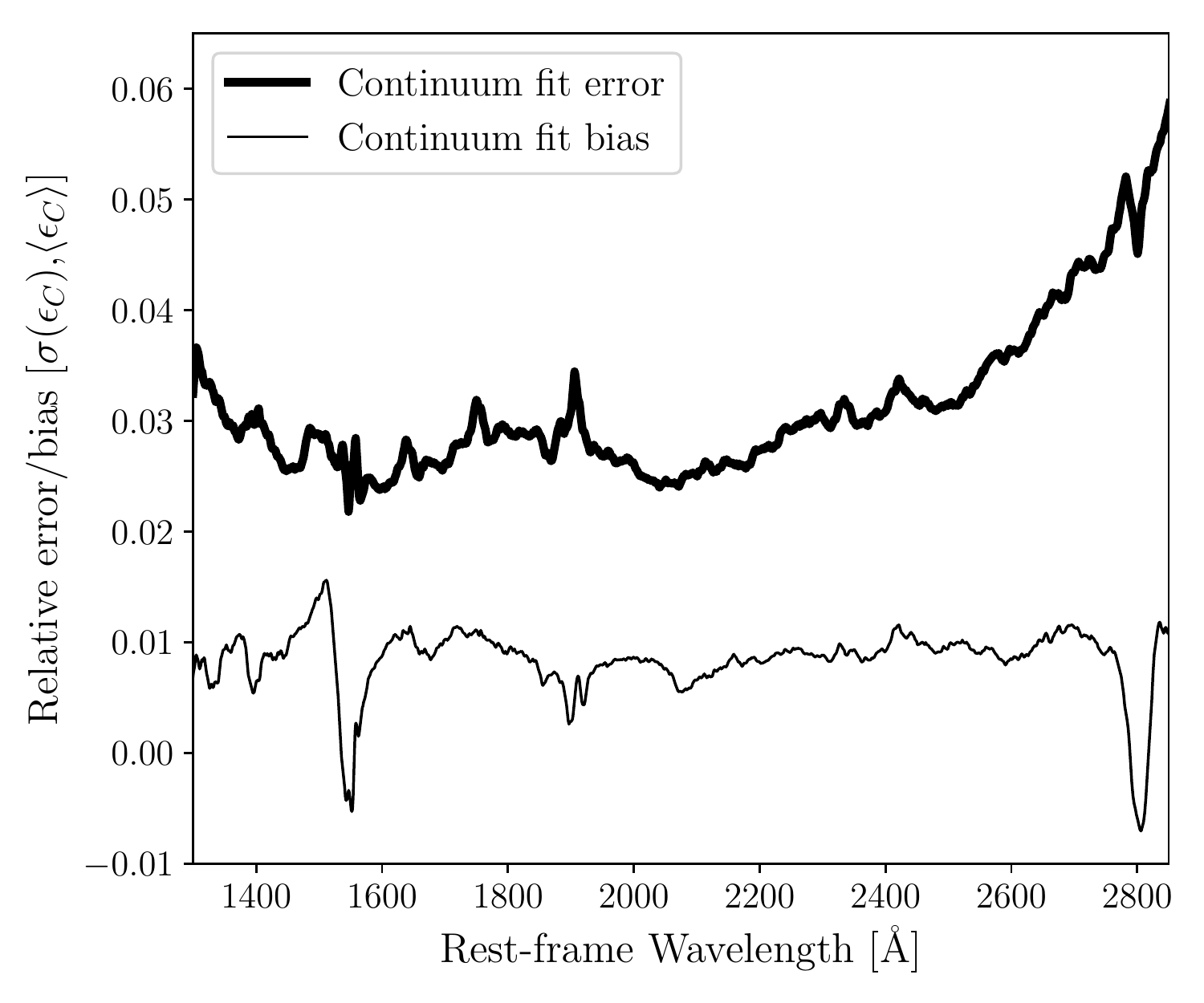}}
\resizebox{8.80cm}{!}{\includegraphics[trim={1.0em 1.0em 5.2em 0}]{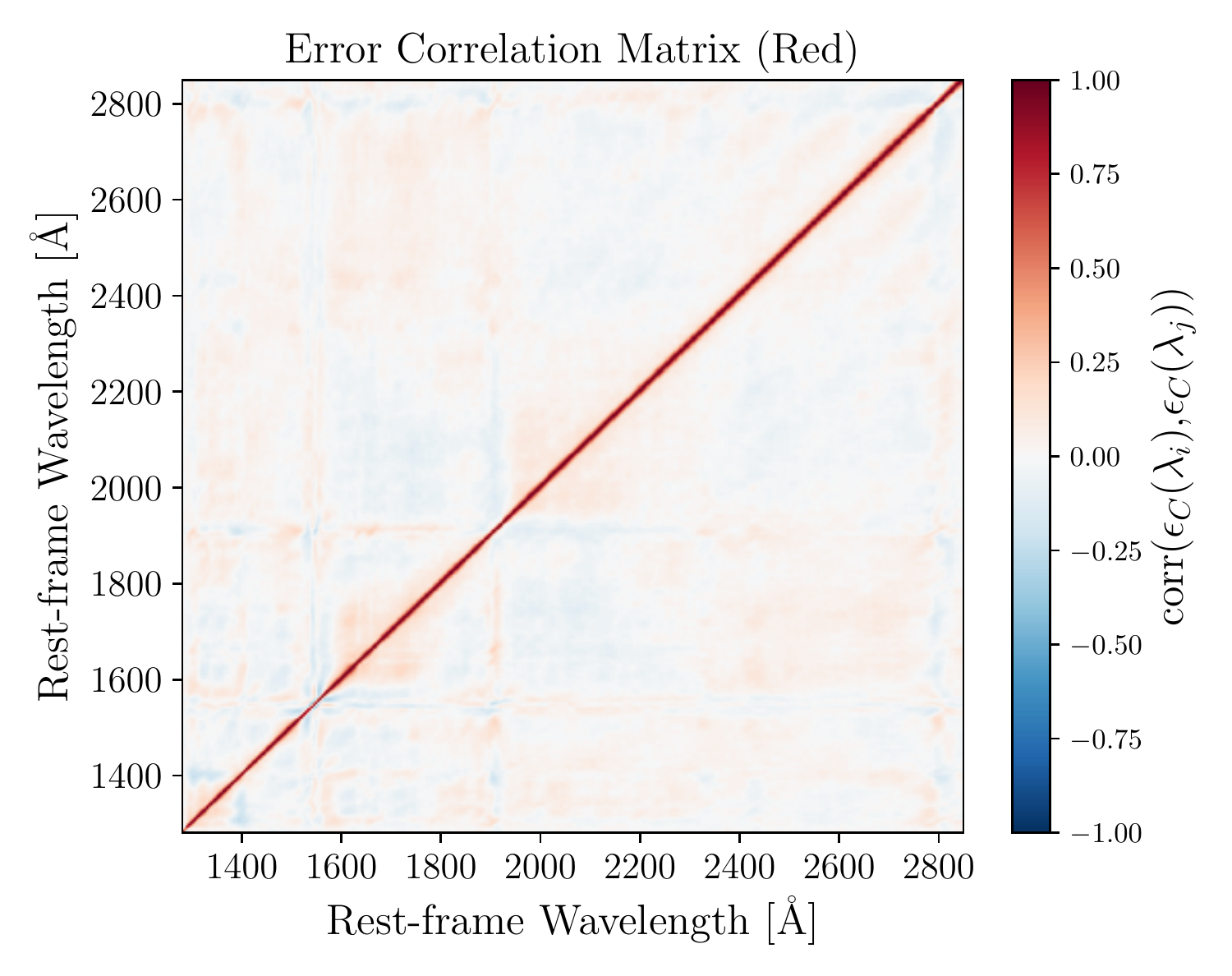}}
\end{center}
\caption{Left: Relative error ($1\sigma$, thick curve) and mean bias (thin curve) of the red-side PCA continuum fits of the full training set sample. Right: Correlation matrix of red-side PCA continuum fit errors.}
\label{fig:fit_err_red}
\end{figure*}

\section{Quantifying Uncertainties in the Continuum Predictions}\label{sec:error}

There are several sources of uncertainty in the prediction of the blue-side continuum, including stochasticity in the relationship between red-side and blue-side features and the inability of the PCA model to exactly reproduce a given spectrum. We quantify the sum of these uncertainties by testing the full predictive procedure on every quasar in the training set, computing the relative continuum error $\epsilon_C \equiv |F_{\rm pred}-F_{\rm true}|/F_{\rm pred}$, where $F_{\rm pred}$ is the predicted flux and $F_{\rm true}$ is the ``true" flux.
For the purposes of this analysis we consider the auto-fit continuum of each quasar to be the ``true" continuum, although this introduces an additional source of noise, and a source of uncertainty in the actual true continuum that we do not attempt to quantify here. From simulations of mock Ly$\alpha$ forest absorption applied to noisy spectra at roughly half the resolution of the SDSS/BOSS spectra used here, \citet{Dall'Aglio09} found that the auto-fit continua were biased low by a few percent in the Ly$\alpha$ forest at $z\sim2$--$2.5$. In this work we ignore this bias because the Ly$\alpha$ forest bias only affects the spectrum blueward of Ly$\alpha$, which is strongly absorbed in high-redshift quasars.

In the left panel of Figure~\ref{fig:fit_err_blue}, we show the mean and standard deviation of $\epsilon_C$ as thin and thick curves, respectively, as a function of rest-frame wavelength (where the rest-frame of each quasar is defined by its best-fit template redshift $z_{\rm temp}$) after three iterations of clipping individual pixels that deviate more than 3 standard deviations from the mean (resulting in $\sim2\%$ of all blue-side pixels masked). The $1\sigma$ error in the region most relevant to damping wing studies, $1210 < \lambda_{\rm rest} < 1250$ {\AA}, is $\sim6$--$12\%$, comparable to the $\sim9\%$ error of the parametric method in \citet{Greig17a}\footnote{\citet{Greig17a} state that $\sim90\%$ of their predicted fluxes at $\lambda_{\rm rest}=1220$ {\AA} lie within $15\%$ of the true continuum, corresponding to $\sim\pm1.64\sigma$ assuming Gaussian distributed errors.}.

Errors in the continuum prediction are strongly correlated across neighboring pixels, in part due to small correlated errors in the smoothed spline fit continua that we assume to be the ``true" continua, but mostly due to smooth variations in the shape of broad emission lines and features of the underlying continuum reconstructed by the PCA model. We show the correlation matrix of $\epsilon_C$ in the right panel of Figure~\ref{fig:fit_err_blue}. The prediction uncertainty is strongly correlated on the scale of the spline fit (the roughly fixed width along the diagonal), and shows larger-scale correlations due to variations in the strengths of broad \ion{N}{5} ($\lambda_{\rm rest}\sim1240$ {\AA}) and \ion{Si}{2} ($\lambda_{\rm rest}\sim1260$ {\AA}) emission lines. These strongly correlated continuum uncertainties, not limited to our method \citep{KH09}, are a critical feature of quasar damping wing analyses that must be fully propagated when conducting parameter inference.

We also note that the red-side fits are not perfect, i.e., the PCA basis is unable to exactly reproduce the input spectra. Assuming again that the auto-fit continuum models represent the ``true" continua, we show the $1\sigma$ relative error and mean bias of the red-side fits in the left panel of Figure~\ref{fig:fit_err_red}. The error on the best-fit red-side PCA continuum is typically $\sim3\%$, increasing smoothly above $\lambda_{\rm rest}\sim2100$ {\AA} to $\sim5\%$ at $\sim2800$ {\AA}, close to the \ion{Mg}{2} broad emission line. Interestingly, the typical red-side fit is biased by $\sim1\%$ across the entire wavelength range, except for small regions close to the peaks of the \ion{C}{4} and \ion{Mg}{2} broad emission lines where the sign is reversed. This bias likely comes about because the actual pixels are used to fit the PCA model of the spectrum instead of the auto-fit continua,
which may be biased slightly high due to differences in how outlier pixels are rejected. We also show the correlation matrix of the red-side fit errors in the right panel of Figure~\ref{fig:fit_err_red}.
  
\begin{figure}[htb]
\begin{center}
\resizebox{8.50cm}{!}{\includegraphics[trim={1em 1.3em 1em 1.0em},clip]{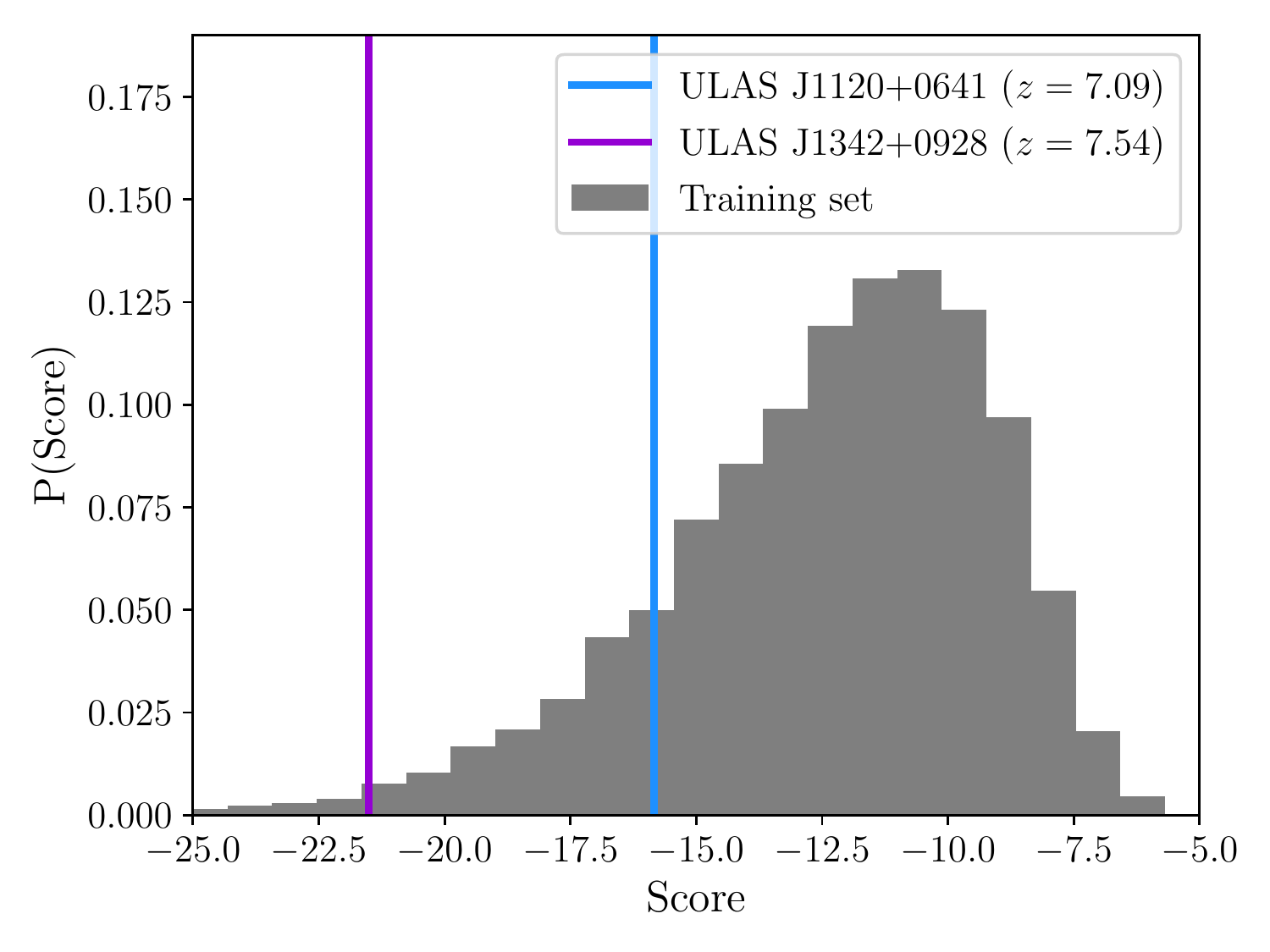}}\\
\end{center}
\caption{Distribution of ``score" -- log of GMM probability (see \S~\ref{sec:z7pred}) -- for the best-fit red-side coefficients of the training set quasars (histogram). Quasars with a high score are located near the peak of the distribution in red-side PCA coefficient space (i.e. ``typical" quasars), while quasars with a low score are located in the outskirts of the distribution. The scores of the two $z>7$ quasars are indicated by vertical lines. Despite the extreme nature of their \ion{C}{4} blueshifts relative to their systemic frames, the two $z>7$ quasars are not extreme outliers of the score distribution, appearing at the $15.0\%$-ile (ULAS J1120+0641) and $1.5\%$-ile (ULAS J1342+0928).}
\label{fig:z7_likely}
\end{figure}
  
\begin{figure*}[h]
\begin{center}
\resizebox{17.6cm}{!}{\includegraphics[trim={8em 1em 8.0em 5.1em},clip]{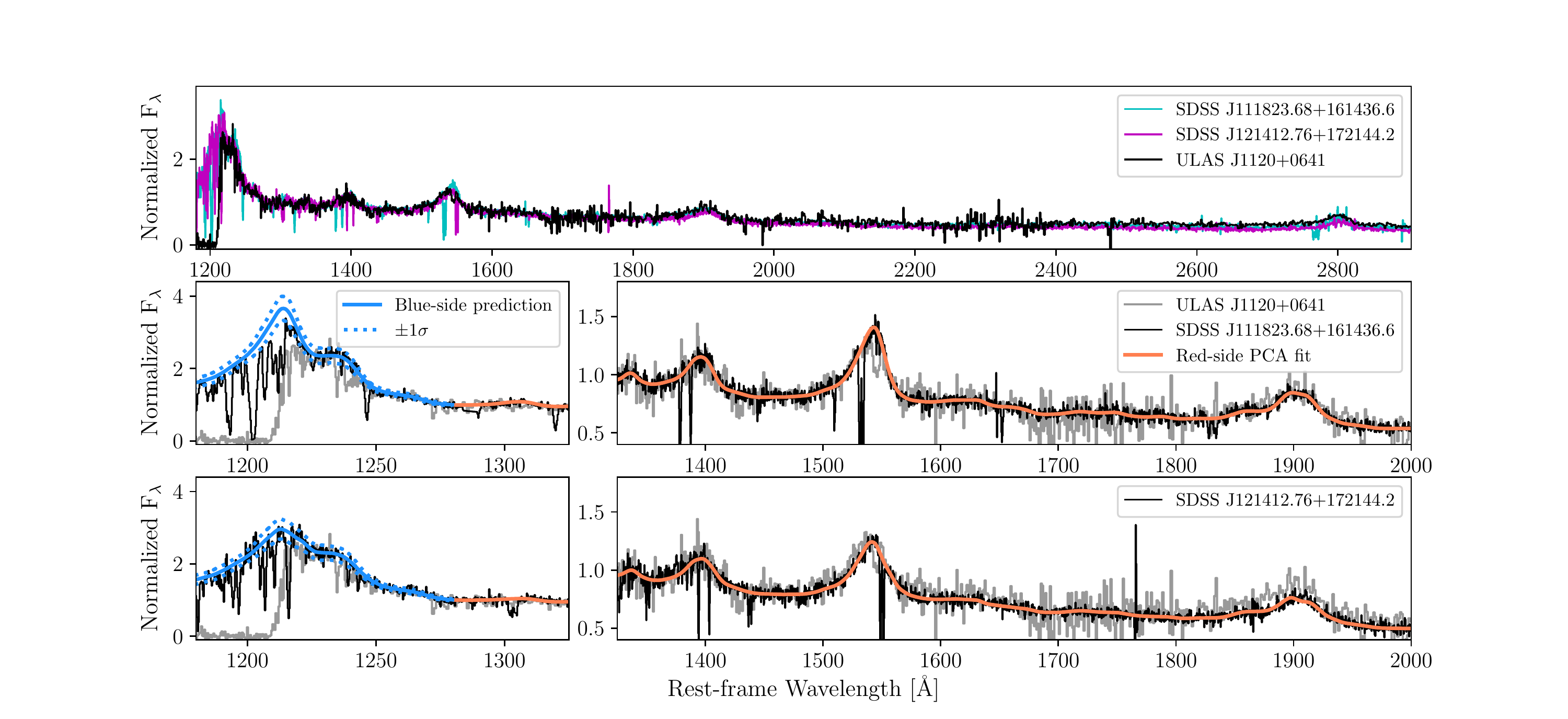}}
\end{center}
\caption{Top: Comparison between the spectrum of ULAS J1120+0641 (black) and its two nearest neighbors in red-side PCA coefficient space: SDSS J111823.68+161436.6 (blue; $D_r=1.24$) and SDSS J121412.76+172144.2 (magenta; $D_r=1.58$). The wavelength axis is presented in the best-fit template redshift frame so that all three spectra shown have a consistently defined redshift. Lower panels: Zoom in on the red-side fit (orange curve; right panels) and blue-side projection (blue curve; left panels) for the nearest neighbor quasars. The dotted blue curves in the left panel show $\pm1\sigma$ continuum prediction uncertainty.}
\label{fig:j1120_nn}
\end{figure*}

\begin{figure*}[h]
\begin{center}
\resizebox{17.6cm}{!}{\includegraphics[trim={8em 1em 8.0em 5.1em},clip]{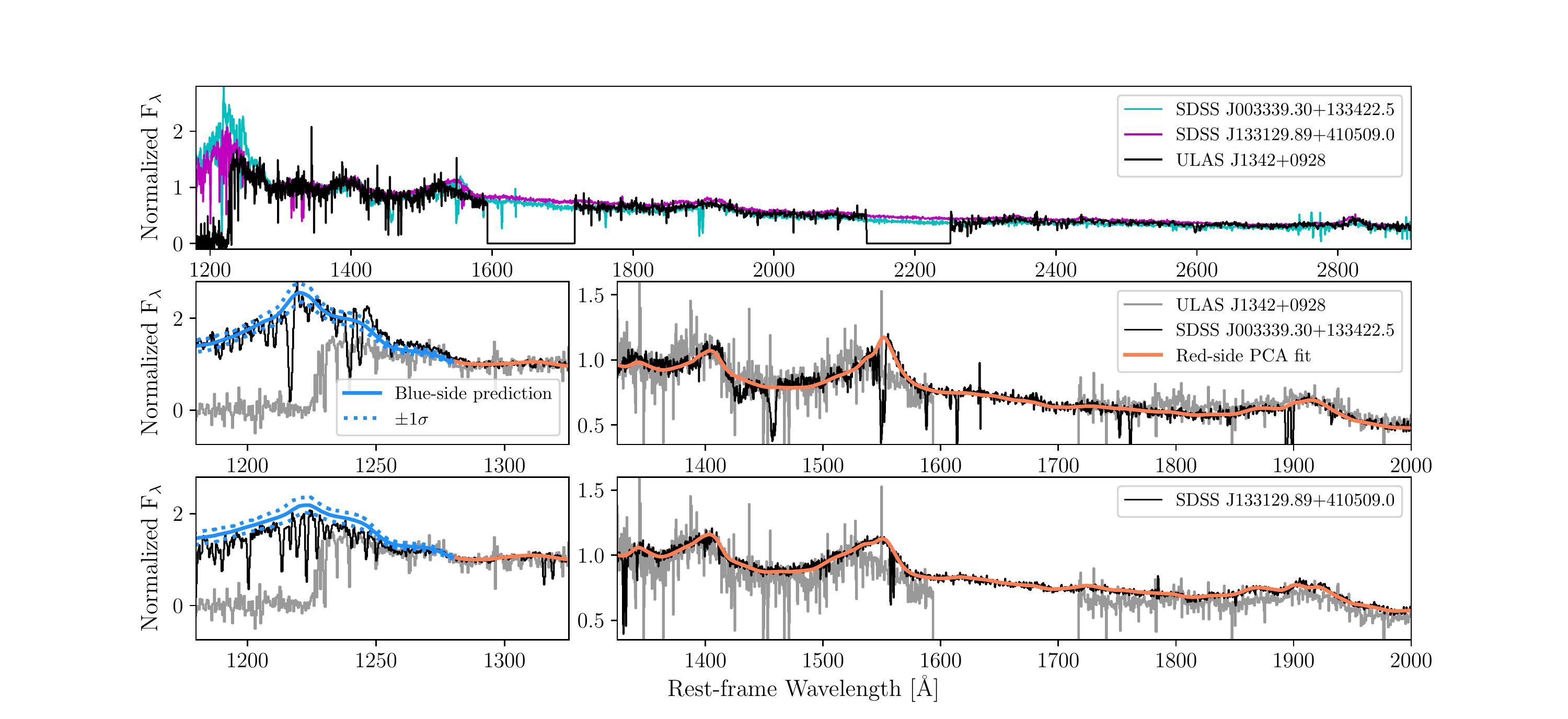}}
\end{center}
\caption{Similar to Figure~\ref{fig:j1120_nn} but for ULAS J1342+0928 and its two nearest neighbors: SDSS J003339.3+133422.5 (blue; $D_r=2.30$) and SDSS J133129.89+410509.0 (magenta; $D_r=2.46$). The substantial difference between systemic and template redshift frames is evident from the onset of saturated Ly$\alpha$ absorption at $\lambda_{\rm rest}\sim1225$ {\AA} in ULAS J1342+0928. While globally the spectra look very similar, the \ion{C}{4} profiles of the nearest neighbor quasars appear to differ somewhat from that of ULAS J1342+0928.}
\label{fig:j1342_nn}
\end{figure*}

\begin{figure*}[htb]
\begin{center}
\resizebox{17.6cm}{!}{\includegraphics[trim={8em 1em 8.0em 5.1em},clip]{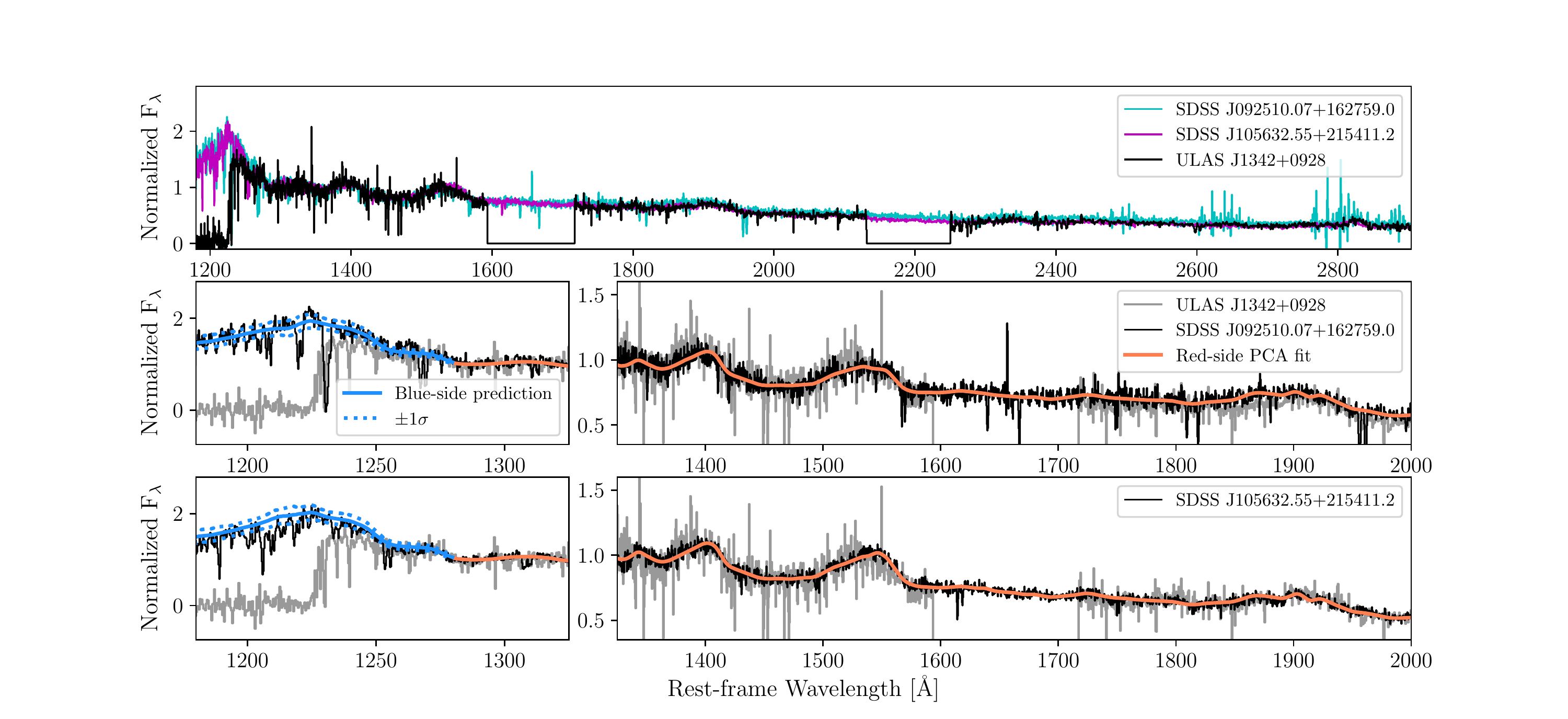}}
\end{center}
\caption{Similar to Figures~\ref{fig:j1120_nn} and \ref{fig:j1342_nn} but for two other nearest neighbor quasars to ULAS J1342+0928: SDSS J092510.07+162759.0 (blue; $D_r=2.70$, 12th-nearest) and SDSS J105632.55+215411.2 (magenta; $D_r=2.92$, 29th-nearest) which have been chosen by eye to have more similar \ion{C}{4} line profiles than the neighbors shown in Figure~\ref{fig:j1342_nn}.}
\label{fig:j1342_nn2}
\end{figure*}

\begin{figure}[htb]
\begin{center}
\resizebox{8.50cm}{!}{\includegraphics[trim={1em 1.2em 1em 0.5em},clip]{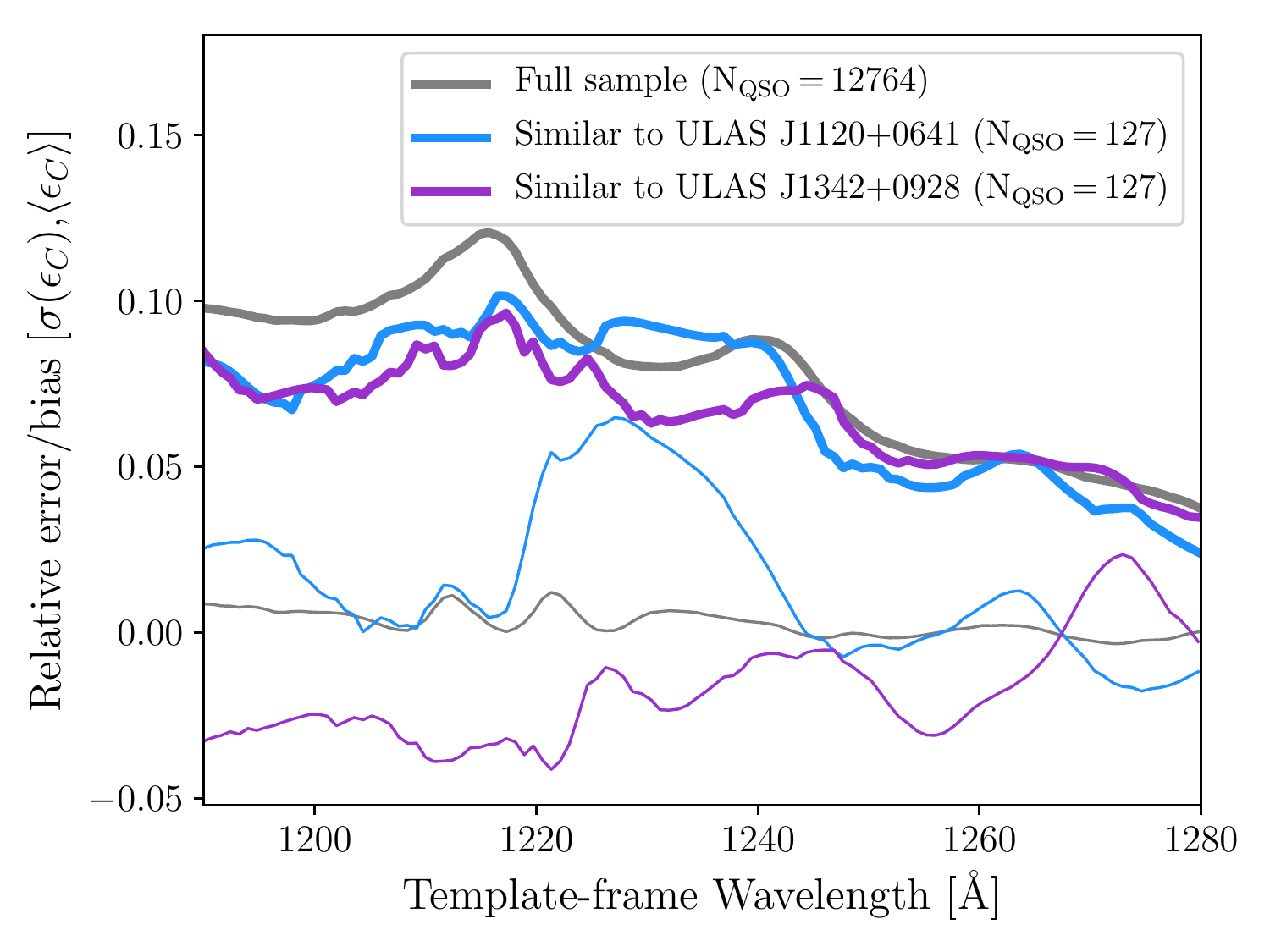}}\\
\end{center}
\caption{Relative uncertainty ($1\sigma$, thick curves) and mean bias (thin curves) for the $1\%$ of quasars in the training set with red-side coefficients most similar to ULAS J1120+0641 (blue) and ULAS J1342+0928 (purple), compared to the full training set sample (grey).}
\label{fig:z7_err}
\end{figure}

\section{Predicted Blue-Side Continua of the Known $z>7$ Quasars}\label{sec:z7pred}

We now demonstrate our continuum fitting machinery on the two highest-redshift quasars known: ULAS J1120+0641 ($z=7.09$, \citealt{Mortlock11}) and ULAS J1342+0928 ($z=7.54$, \citealt{Banados17}). Modeling the intrinsic continuum of these quasars, and understanding the uncertainty in those models, is critical to constraining the neutral fraction of the IGM, so they provide a first test of the applicability of our continuum modeling procedure.

For our analysis of the two $z>7$ quasars, we use the previously published spectra from their respective discovery papers. We use the combined VLT/FORS2 + Gemini/GNIRS spectrum of ULAS J1120+0641 published in \citet{Mortlock11} and we use the combined Magellan/FIRE + Gemini/GNIRS spectrum of ULAS J1342+0928 published in \citet{Banados17}. We fit the red-side continua of these quasars nearly identically to our fits of the training set, performing $\chi^2$ minimization to fit both their red-side coefficients $r_i$ and their template redshifts $z_{\rm temp}$. As with the training set, we reject pixels that deviate from an automated spline fit of the continuum by $>3\sigma$ to reject strong metal absorption lines. In addition, we mask spectral regions corresponding to regions of poor atmospheric transmission in the gaps between the J/H/K bands, and for ULAS J1120+0641 we mask regions of strong metal line absorption reported by \citet{Bosman17} from their extremely deep VLT/X-Shooter spectrum. For ULAS J1120+0641 we find $z_{\rm temp}=7.0834$, a very small blueshift of ${\Delta}v=63$ km/s from the systemic redshift ($z_{\rm sys}=7.0851$ from host galaxy [\ion{C}{2}] emission, \citealt{Venemans17}), while for ULAS J1342+0928 we find $z_{\rm temp}=7.4438$, a blueshift of ${\Delta}v=3422$ km/s from the systemic redshift ($z_{\rm sys}=7.5413$ from host galaxy [\ion{C}{2}] emission, \citealt{Venemans17b}).

Both of these quasars have peculiar broad emission line properties, most notably large blueshifts in the \ion{C}{4} line relative to \ion{Mg}{2}: ${\Delta v}\sim2800$ km/s for ULAS J1120+0641, and ${\Delta v}\sim6100$ km/s for ULAS J1342+0928. We quantify the outlying nature of these quasars in the context of our PCA model by modeling the 10-dimensional probability distribution of the best-fit $r_i$ from the training set as a mixture of multivariate Gaussians, i.e. a Gaussian mixture model (GMM),
 using the \textsc{GaussianMixture} package in \textsc{scikit-learn}. We chose the number of Gaussians $N_{\rm gauss}=9$ to minimize the Bayesian information criterion (BIC; \citealt{Schwarz78}), defined by 
\begin{equation}
{\rm BIC} = k\log{N_{\rm spec}} - 2\log{\hat{L}},
\end{equation}
where $k$ is the number of model parameters of the GMM (i.e. means, amplitudes, and covariances of the individual Gaussians), 
\begin{equation}
k=N_{\rm gauss}\times\left(\frac{1}{2}N_{\rm PCA}^2+\frac{3}{2}N_{\rm PCA}+1\right),
\end{equation}
$N_{\rm spec}$ is the number of spectra in the training set, and $\hat{L}$ is the maximum likelihood (i.e. the product of the GMM evaluated for each quasar) for the given number of Gaussians. The distribution of ``scores" of each quasar in the training set, defined as the log of the GMM probability evaluated at their corresponding values of $r_i$, is shown in Figure~\ref{fig:z7_likely}. Quasars whose spectra have a higher score are located closer to the mean quasar spectrum, while lower scores represent outliers. The scores of ULAS J1120+0641 and ULAS J1342+0928 are shown by the vertical blue and purple lines corresponding to percentiles of $\sim15.0\%$ and $1.5\%$ in the distribution of scores, respectively. Given the extreme broad emission line properties of these two quasars \citep{Mortlock11,BB15,Banados17}, these percentiles may seem somewhat high -- however, the GMM score is sensitive to more than just the particularly extreme features of the $z>7$ quasar spectra (e.g. their large \ion{C}{4} blueshifts), illustrating that in other aspects the quasars are more representative of the bulk sample at low redshift (e.g. continuum slope and equivalent widths of broad emission lines).

Our ability to constrain the intrinsic blue-side continua of peculiar quasars like ULAS J1120+0641 and ULAS J1342+0928 can be determined by testing how well we can make predictions for similar quasars, i.e. with similar $r_i$, in the training set. We define a distance in $r_i$ space by
\begin{equation}\label{eqn:dist}
D_r \equiv \sqrt{\sum_i^{N_{\rm PCA,r}}\left(\frac{{\Delta}r_i}{\sigma(r_i)}\right)^2},
\end{equation}
where $i$ is the PCA component index, $N_{\rm PCA,r}=10$ is the number of red-side PCA basis vectors, ${\Delta}r_i$ is the difference between the best-fit $r_i$ values, and $\sigma(r_i)$ is the standard deviation of best-fit $r_i$ values in the training set sample. The median distance between randomly-chosen quasars is $D_{r,\rm med} \sim7.5$. We then identify the $1\%$ of quasars ($N_{\rm QSO}=127$) in the training set with the smallest $D_r$ to the quasar whose continuum we are predicting, i.e. the set of nearest neighbor quasar spectra in the training set. In Figures~\ref{fig:j1120_nn} and \ref{fig:j1342_nn} we show the two nearest neighbors (i.e. the first and second smallest $D_r$) to ULAS J1120+0641 and ULAS J1342+0928, respectively, along with their respective red-side PCA fits and blue-side predicted continua.  While the ULAS J1120+0641 neighbors have strikingly similar red-side broad emission line profiles, the ULAS J1342+0928 neighbors are less similar, reflecting the more sparse sampling of the distribution of quasar spectra. More qualitatively similar spectra exist in the set of nearest neighbors, however, and we show two examples in Figure~\ref{fig:j1342_nn2}, which were chosen by eye solely from a comparison of their red-side spectrum to that of ULAS J1342+0928.

The blue and purple curves in Figure~\ref{fig:z7_err} show the relative error and mean bias for the ULAS J1120+0641 and ULAS J1342+0928 nearest neighbor samples, respectively, as a function of rest-frame ($z_{\rm temp}$-frame) wavelength.
Quasars similar to ULAS J1342+0928 tend to have smaller continuum errors than average at $\lambda_{\rm rest}<1245$ {\AA} with a modest bias, while those similar to ULAS J1120+0641 tend to have error comparable to the full training set and a $\sim5\%$ bias at $1220$ {\AA} $\la\lambda_{\rm rest}\la1235$ {\AA}. In Figures \ref{fig:j1120_nn} through \ref{fig:j1342_nn2} we have corrected the blue-side predictions for their corresponding mean biases, and we similarly correct the predictions for the $z>7$ quasars shown below.

To model the relative uncertainties between our blue-side predictions and the observed spectra, we approximate the continuum error distribution as a multivariate Gaussian distribution with mean and covariance set by the mean and covariance matrix of prediction errors $\epsilon_C$ measured from $127$ nearest neighbor quasars as described above.
To generate plausible representations of the true continuum $F_{\rm sample}$, we draw samples of $\epsilon_C$ from this multivariate Gaussian and multiply the blue-side prediction by $1+\epsilon_C$, i.e. $F_{\rm sample} = F_{\rm pred} \times (1+\epsilon_C)$. This procedure can be used to draw Monte Carlo samples of the uncertainty in the continuum prediction for analyses of, e.g., the damping wing from the IGM.

\begin{figure*}[htb]
\begin{center}
\resizebox{17.6cm}{!}{\includegraphics[trim={7em 2.5em 8.0em 6.8em},clip]{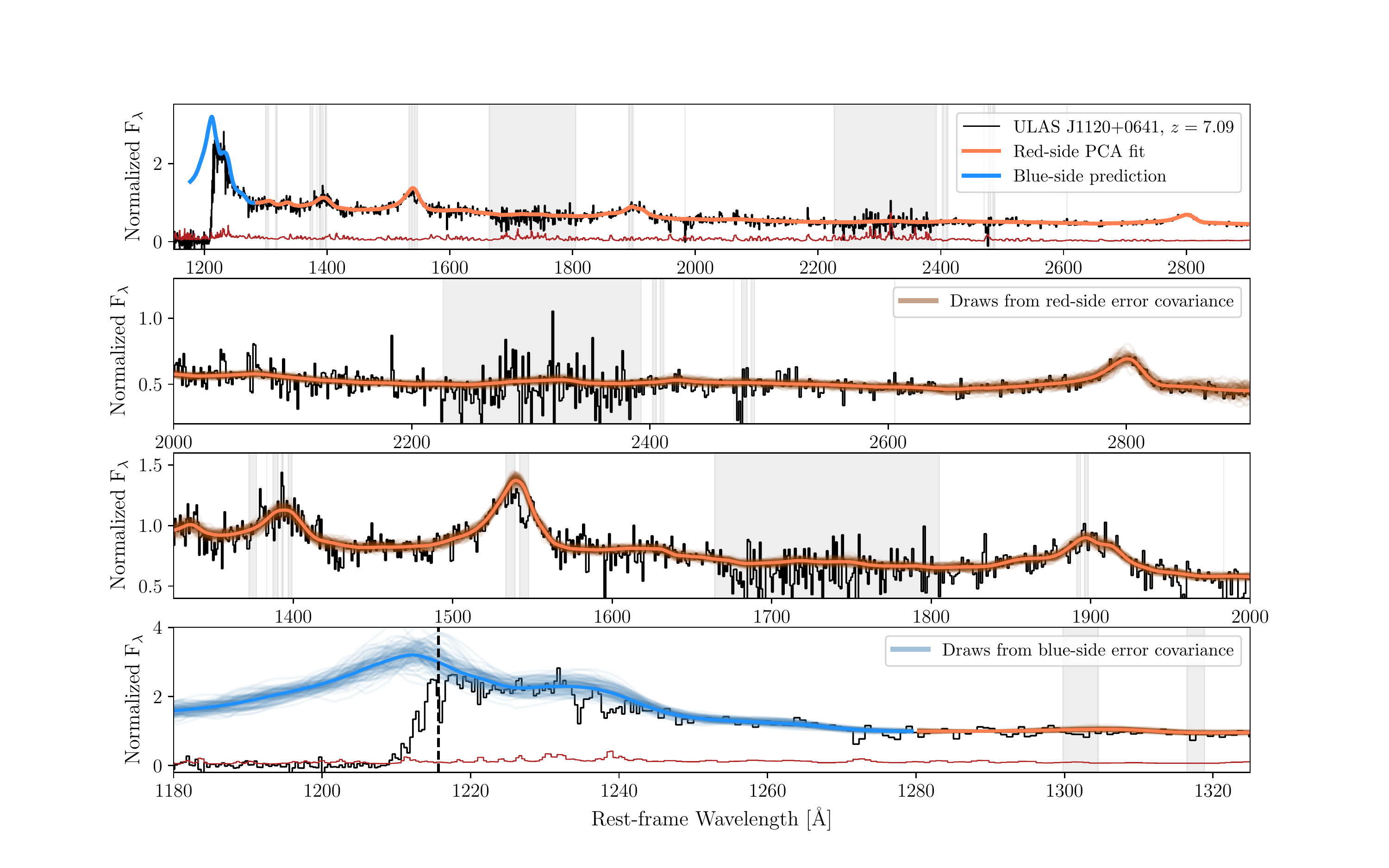}}\\
\end{center}
\vspace{-1em}
\caption{Top: FORS2+GNIRS spectrum of ULAS J1120+0641 (black) and its noise vector (red) from \citet{Mortlock11}. The red-side PCA fit and bias-corrected blue-side prediction are shown as the orange and blue curves, respectively. Grey shaded regions represent pixels that were masked out when performing the red-side fit. Middle: Zoom in on the red-side fit of the strongest broad emission lines. The brown transparent curves show 100 draws from the covariant red-side fit error shown in Figure~\ref{fig:fit_err_red}. Bottom: Zoom in of the blue-side spectrum, where the vertical dashed line corresponds to rest-frame Ly$\alpha$. The blue transparent curves show 100 draws from the covariant blue-side prediction error measured for 127 similar quasars in the training set.}
\label{fig:mortlock_pca}
\end{figure*}

\begin{figure*}[htb]
\begin{center}
\resizebox{17.6cm}{!}{\includegraphics[trim={7em 2.5em 8.0em 6.8em},clip]{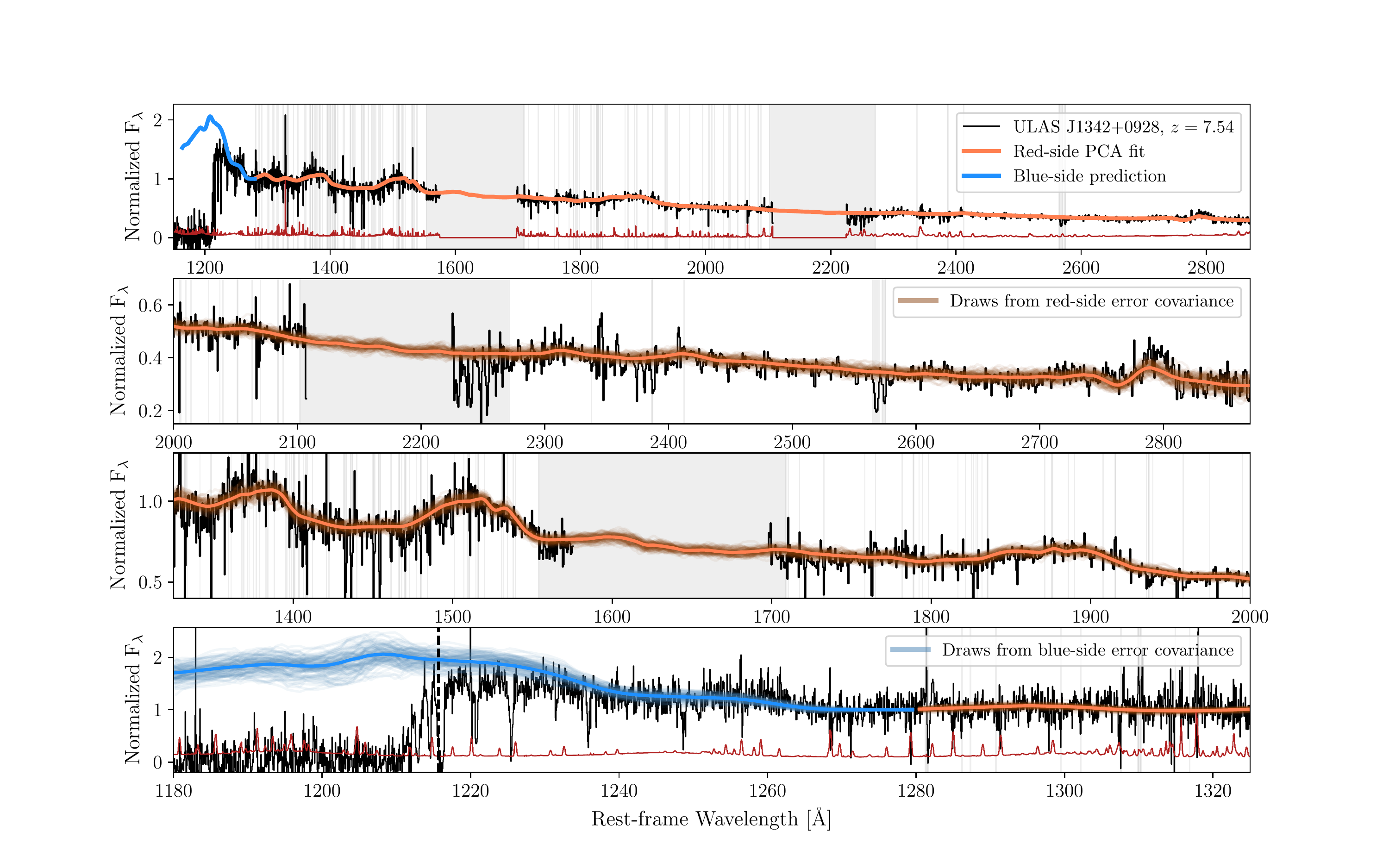}}\\
\end{center}
\vspace{-1em}
\caption{Similar to Figure~\ref{fig:mortlock_pca} but for the FIRE+GNIRS spectrum of ULAS J1342+0928 from \citet{Banados17}. The FIRE portion of the spectrum in the top two panels has been re-binned to match the pixel scale of the GNIRS data used in the K-band, while the bottom panel is shown at the original FIRE pixel scale to better highlight .}
\label{fig:pisco_pca}
\end{figure*}

In Figure~\ref{fig:mortlock_pca} we show the red-side continuum fit (orange) and blue-side prediction (blue) for ULAS J1120+0641, where the latter has been corrected for the mean bias of the prediction for similar quasars (i.e. the thin blue curve in Figure~\ref{fig:z7_err}). The pixel mask for the red-side fit is shown by the grey shaded regions. As shown in the middle panels, the PCA is able to fit the \ion{Si}{4}, \ion{C}{4}, \ion{C}{3}], and \ion{Mg}{2} broad emission lines very well, although most of the core of the \ion{C}{4} line has been masked out due to associated \ion{C}{4} absorbers which are unresolved in the GNIRS data. The blue-side prediction, shown more closely in the bottom panel, matches the blue-side continuum very closely for $\lambda_{\rm rest}>1225$ {\AA}, but at $\lambda_{\rm rest}\sim1216$--$1225$ {\AA} there is evidence that the observed spectrum lies below the predicted continuum. This deficit suggests the presence of an IGM damping wing, as reported by \citet{Mortlock11} and \citet{Greig17b}. However, from the covariant error draws $F_{\rm sample}$, shown as the transparent curves, it is clear that the uncertainty in our continuum model is of comparable amplitude to any putative damping wing signal, so any reionization constraint based solely on this damping wing signature will be weak at best.

In Figure~\ref{fig:pisco_pca} we show the red-side continuum fit and bias-corrected blue-side prediction for ULAS J1342+0928, analogous to Figure~\ref{fig:mortlock_pca}. The middle panels show that the PCA model is able to fit the red-side spectrum reasonably well despite the odd shape of the broad emission lines, although the amplitude of the \ion{Mg}{2} line is somewhat weaker in the fit compared to the actual spectrum. Contrary to ULAS J1120+0641, and in agreement with the analysis in \citet{Banados17} based on a matched composite spectrum, the ULAS J1342+0928 spectrum shows a significant, extended deficit relative to the blue-side prediction. This deficit is too large to explain with continuum error, and the deficit smoothly increases from $\lambda_{\rm rest}\sim1250$ {\AA} to $\lambda_{\rm Ly\alpha}$ in qualitative agreement with the expected profile of a Ly$\alpha$ damping wing from a substantially neutral IGM.

We will determine quantitative constraints on the reionization epoch and quasar lifetimes from the continuum-normalized spectra of ULAS J1120+0641 and ULAS J1342+0928 in a subsequent paper (Davies et al., in prep.).

\section{Conclusion}

In this work, we have developed a PCA-based method for predicting the intrinsic quasar continuum at rest-frame wavelengths $\lambda_{\rm rest}<1280$ {\AA} from the properties of the spectrum at $1280$ {\AA} $ < \lambda_{\rm rest} < 2850$ {\AA}. We exploited the large number of high-quality quasar spectra from SDSS/BOSS whose broad wavelength coverage enables building a continuous spectral model covering $1175$ {\AA} $ < \lambda_{\rm rest} < 2900$ {\AA} from a sample of $12,764$ spectra with ${\rm S\slash N}>7$. After initial processing of the spectra with adaptive, piecewise spline fits and subsequent nearest-neighbor stacking, we performed a log-space PCA decomposition of the training set truncated at 10 red-side and 6 blue-side basis spectra. We determined the best-fit values of these coefficients, and red-side template redshifts, for each quasar spectrum in the training set, and derived a projection matrix relating the red-side and blue-side coefficients. This projection matrix can then be used to predict the blue-side coefficients (and thus the blue-side spectrum) from a fit to the red-side coefficients (and template redshift) of an arbitrary quasar spectrum.

By testing our procedure on the training set, we found that we can predict the blue-side spectrum of an individual quasar to $\sim6$--$12\%$ precision with very little mean bias ($\la1\%$), although prediction errors are strongly covariant across the entire blue-side spectrum. As a proof-of-concept test, we predicted the blue-side spectra of two $z>7$ quasars thought to exhibit damping wings due to neutral hydrogen in the IGM, ULAS J1120+0641 \citep{Mortlock11} and ULAS J1342+0928 \citep{Banados17}. These two quasars are known to possess outlying spectral features from the primary locus of quasar spectra at lower redshift, so we established that our method works similarly well on such outliers by testing the machinery on the $1\%$ nearest neighbor quasars in the training set to each $z>7$ quasar. While ULAS J1120+0641 shows only modest evidence for an IGM damping wing, ULAS J1342+0928 appears to be strongly absorbed redward of systemic-frame Ly$\alpha$. In a subsequent paper we will constrain the neutral fraction of the IGM at $z>7$ through statistical analysis of these two spectra.

Our relatively unbiased method for quasar continuum prediction can be applied more broadly to quasar proximity zones at any redshift where direct measurement of the continuum is difficult, i.e. $z>4$. Measurements of the quasar proximity effect using our predicted continua can in principle constrain the strength of the ionizing background (e.g. \citealt{Dall'Aglio08}), the helium reionization history from the ``thermal" proximity effect (\citealt{Khrykin17}, Hennawi et al., in prep.), and timescales of quasar activity \citep{Eilers17}. The predicted continua may also be useful for analyzing proximate absorption systems such as damped Ly$\alpha$ absorbers at $z\ga5$.

\section*{Acknowledgements}

We would like to thank D. Stern for supporting the discovery of ULAS J1342+0928, and M. Turner for consultation and support in the early efforts to model the intrinsic continuum of ULAS J1342+0928. 

EPF, BPV, and FW acknowledge funding through the ERC grant `Cosmic Dawn'.

Funding for SDSS-III has been provided by the Alfred P. Sloan Foundation, the Participating Institutions, the National Science Foundation, and the U.S. Department of Energy Office of Science. The SDSS-III web site is http://www.sdss3.org/.

SDSS-III is managed by the Astrophysical Research Consortium for the Participating Institutions of the SDSS-III Collaboration including the University of Arizona, the Brazilian Participation Group, Brookhaven National Laboratory, Carnegie Mellon University, University of Florida, the French Participation Group, the German Participation Group, Harvard University, the Instituto de Astrofisica de Canarias, the Michigan State/Notre Dame/JINA Participation Group, Johns Hopkins University, Lawrence Berkeley National Laboratory, Max Planck Institute for Astrophysics, Max Planck Institute for Extraterrestrial Physics, New Mexico State University, New York University, Ohio State University, Pennsylvania State University, University of Portsmouth, Princeton University, the Spanish Participation Group, University of Tokyo, University of Utah, Vanderbilt University, University of Virginia, University of Washington, and Yale University.

 \newcommand{\noop}[1]{}

\end{document}